\documentclass[journal]{IEEEtran}
\usepackage{cite}
\usepackage{graphicx}
\usepackage{amsmath}
\usepackage{array}
\usepackage{mdwmath}
\usepackage{amssymb}
\usepackage{mdwtab}
\usepackage{url}
\usepackage{stfloats}
\usepackage{amsmath,amsthm}
\usepackage{verbatim}
\usepackage{threeparttable}
\usepackage{color}
\usepackage{xcolor}
\usepackage{bm}
\usepackage[normalem]{ulem}

\newtheorem{theorem}{Theorem}

\newcommand{\tabincell}[2]{\begin{tabular}{@{}#1@{}}#2\end{tabular}} 
\usepackage{booktabs}

\ifCLASSOPTIONcompsoc
\usepackage[caption=false,font=normalsize,labelfont=sf,textfont=sf]{subfig}
\else
\usepackage[caption=false,font=footnotesize]{subfig}
\fi

\allowdisplaybreaks[2]

\everymath{\displaystyle}

\begin{document}

\title{Multi-Task Over-the-Air Federated Learning in Cell-Free Massive MIMO Systems}

\author{Chen~Chen, \textit{Member, IEEE,}
Emil Björnson, \textit{Fellow, IEEE,}
Carlo Fischione, \textit{Fellow, IEEE}

\thanks{
Manuscript received xxx; revised xxx; accepted xxx. Date of publication xxx; date of current version xxx.
This work was supported by the KTH Digital Future research center. 
An earlier version of this paper was
presented in part at the 2024 IEEE International Workshop on Computer Aided Modeling and Design of Communication Links and Networks (IEEE CAMAD 2024)~\cite{chen2024over}.

The authors are with the School of Electrical Engineering and Computer Science, KTH Royal Institute of Technology, Stockholm, Sweden (e-mail: \{chch2, emilbjo, carlofi\}@kth.se).
}

\thanks{Color versions of one or more of the figures in this paper are available online at http://ieeexplore.ieee.org.}
\thanks{Digital Object Identifier xxx}
}

\markboth{IEEE}%
{Shell \MakeLowercase{\textit{et al.}}: Bare Demo of IEEEtran.cls for Journals}
\maketitle

\begin{abstract}
Wireless devices are expected to provide a wide range of AI services in 6G networks.
The increasing computing capabilities of wireless devices and the surge of wireless data motivate the use of privacy-preserving federated learning~(FL). In contrast to  centralized learning that requires sending large amounts of raw data during uplink transmission, only local model parameters are uploaded in FL. 
Meanwhile, over-the-air (OtA) computation  is considered as a communication-efficient solution for fast FL model aggregation by exploiting the superposition properties of wireless multi-access channels. 
The required communication resources in OtA FL do not scale with the number of FL devices.
However, OtA FL is significantly affected by the uneven signal attenuation experienced by different FL devices. 
Moreover, the coexistence of multiple FL groups with different FL tasks brings about inter-group interference. 
These challenges cannot be well addressed by conventional cellular network architectures.
Recently, Cell-free Massive MIMO (mMIMO)
has emerged as a promising technology to provide uniform coverage and high rates via joint coherent transmission.
In this paper, we investigate multi-task OtA FL in Cell-free mMIMO systems. 
We propose optimal designs of transmit coefficients and receive combining at different levels of cooperation among the access points, aiming to minimize the sum of OtA model aggregation errors across all FL groups. 
Numerical results demonstrate that Cell-free mMIMO significantly outperforms conventional Cellular  mMIMO in term of the FL convergence performance by operating at appropriate cooperation levels.
\end{abstract}


\section{Introduction}
Artificial intelligence (AI) is expected to play an important role in many use-cases of Sixth Generation (6G) networks, such as smart cities, smart grids, autonomous driving, and health monitoring~\cite{Survey1, hellstrom2022wireless}. Meanwhile, the exponential growth of wireless data generated at edge devices promotes the 
connected intelligence.
Conventional centralized learning requires sending a large amount of raw data from distributed wireless devices to a parameter server, e.g., base station (BS). However, transmitting raw data over wireless networks face challenges such as limited communication bandwidth, high communication latency, and security and privacy issues~\cite{Survey2, chen2023secret}. To address these challenges, federated learning (FL) has recently been proposed to enable training a common global AI model in a distributed manner~\cite{kairouz2021advances}.
During each FL training round, wireless devices only transmit local model updates to the parameter server, rather than raw data. The local model updates are aggregated at the parameter server to update the global model, which is then broadcast to the wireless devices. Since only 
AL model parameters are exchanged between wireless devices and the parameter server, FL effectively alleviates the risk of data leakage and protects user privacy.

Albeit FL avoids transmitting large amounts of raw data, the exchange of high-dimensional model parameters over resource-constrained wireless networks remains a challenge. 
There have been research efforts exploring 
wireless resource allocation in conventional digital transmissions to improve the convergence performance of FL~\cite{yang2019scheduling, chen2020joint, chen2020convergence, vu2020cell, jeon2020compressive}. In \cite{yang2019scheduling}, the authors compared different user scheduling policies with regard to the convergence rate of FL.
Joint user scheduling and resource block (RB) allocation were optimized to minimize the FL loss function~\cite{chen2020joint} and convergence time \cite{chen2020convergence}.
Multiple-input multiple-output (MIMO) has been a key technology in today's 5G communication systems~\cite{bjornson2024towards}.
A single BS  equipped with multiple antennas served as a parameter server in~\cite{vu2020cell, jeon2020compressive}  to enhance the communication efficiency and FL convergence performance. 
In~\cite{Afsaneh2024}, adaptive quantization resolution and power control were investigated for FL over Cell-free networks.
In digital transmissions,  the parameter server decodes the signals transmitted by the FL devices individually, and thus there is inter-user interference.
As a consequence,  the main bottleneck of UL uploading of model parameters using digital transmissions is that the required communication resources scale with the number of FL devices. 

An alternative solution is to utilize over-the-air (OtA) computation for fast uplink model  aggregation, which is referred to as  OtA FL.
OtA computation is a resource-efficient method that leverages the waveform superposition property of the multi-access channel to realize fast wireless data aggregation~\cite{csahin2023survey, chen2023over}. 
Different from conventional digital transmissions,
OtA computation is an analog communication scheme that enables a large number of wireless devices to communicate in an interference-free manner using the same time-frequency resource. In other words, the  communication resources required by OtA computation do not increase with the number of wireless devices, which is in sharp contrast with conventional digital transmissions using orthogonal wireless resource allocation.
Instead of decoding individual signals transmitted by the wireless devices, OtA computation aims to reconstruct a function of the transmitted signals. 
Theoretically, OtA computation can reconstruct any nomographic function~\cite{goldenbaum2011analyzing}. 
In OtA FL, the objective of uplink model aggregation is to reconstruct the weighted sum of the local model parameters.

The performance of OtA computation is measured by the distortion between the reconstructed function and the desired function, which is quantified by the mean squared error (MSE). 
Significant efforts have been recently devoted to minimizing the MSE~\cite{cao2020optimized, hellstrom2023federated, wen2019reduced, jing2023transceiver}. 
The authors in~\cite{cao2020optimized} jointly optimized the transmit powers at the wireless devices and the denoising factor  at the receiver.
The work in~\cite{hellstrom2023federated} introduced a retransmission mechanism and optimized the power control over multiple OtA computation transmissions. 
The work in~\cite{wen2019reduced} considered OtA computation in a MIMO system and employed zero-forcing based transmit beamforming.
In~\cite{jing2023transceiver}, the authors  developed the optimal 
transceiver beamforming design for multi-antenna transmitters and receivers. 
It has been revealed that FL can achieve better convergence performance for a smaller OtA model aggregation error, which is measured by the MSE between the OtA aggregated model parameters and the desired model parameters~\cite{asaad2024joint, liu2021reconfigurable, ni2021over, zhang2024federated}. 
In order to minimize the OtA model aggregation error, joint antenna selection and beamforming design were addressed in~\cite{asaad2024joint}, and a passive/active reconfigurable intelligent surface was introduced and configured  in~\cite{liu2021reconfigurable, ni2021over, zhang2024federated}. 

The main challenge facing OtA computation/FL is the computation errors caused by uneven signal attenuation, which is also referred to as the straggler problem. The OtA computation/FL performance is typically limited by the stragglers (wireless devices with the worst channel conditions)~\cite{yang2020federated}.
Unfortunately, this challenge
cannot be well addressed by conventional cellular network architectures, even if multiple antennas are deployed on each parameter server.  
Recently, Cell-free Massive MIMO (mMIMO) has shown great potential to provide ubiquitous and uniform coverage with high spectral and energy efficiency~\cite{ammar2021user, 7827017, buzzi2019user}. The idea is to densely deploy a large number of simple access points (APs) that cooperatively serve all the users in the network; this in turn, eliminates cell boundaries~\cite{bjornson2019making}.
Albeit Cell-free mMIMO has been widely investigated in terms of communication performance, the OtA computation/FL performance has not been well evaluated.
In~\cite{sifaou2023over}, the authors studied OtA  FL in a scalable Cell-free mMIMO network. However, the APs simply adopted suboptimal maximum ratio combining and the FL devices transmitted at full power.
Hence, how to optimally implement OtA computation/FL in Cell-free mMIMO remains largely an open question. 
In practice, there may be multiple FL groups with different FL tasks that interfere with each other. In~\cite{wang2022interference}, the authors considered a multi-cell setup where each cell has a FL group with a specific FL task. Nevertheless, different from cellular communications, a group of FL devices may be distributed across multiple cells. Conventional cellular network architectures may be insufficient to address the large channel variations and strong inter-group interference in multi-task FL.

 Different from the existing work that considers a single FL task~\cite{asaad2024joint, liu2021reconfigurable, ni2021over, zhang2024federated, yang2020federated, sifaou2023over} or multiple FL tasks in a multi-cell network with single-antenna BSs~\cite{wang2022interference},
 in this paper, we investigate the advantages of implementing multi-task OtA FL over Cell-free mMIMO networks.
The major contributions of our work are summarized as follows: 
\begin{itemize}
	\item We study multi-task OtA FL in a Cell-free mMIMO system where different FL groups have different FL tasks. We consider three cooperation levels among the APs and analyze the fronthaul signaling required by them. We derive the closed-form expressions of the OtA model aggregation errors for the considered cooperation levels, taking into account inter-group interference, spatially correlated fading and channel estimation errors.
 \item We derive the upper bound of the gap between the FL training loss and the optimal training loss, and show that the convergence performance can be improved by minimizing the OtA model aggregation error. Based on this communication-learning analysis framework,
 we develop optimal designs of OtA model aggregation designs for the three cooperation levels. More specifically, we jointly optimize the transmit coefficients and receive combining to minimize the weighted sum model aggregation error of multiple FL groups. 
	\item We numerically evaluate the model aggregation errors and FL convergence performance of the proposed OtA model aggregation designs under different device distribution modes and show how Cell-free mMIMO should be operated to outperform conventional Cellular mMIMO.  
\end{itemize}

Our previous work~\cite{chen2024over} jointly optimized the transmit coefficients and  receive combining to minimize the OtA computation error in a Cell-free mMIMO system for different cooperation levels among the APs. In this paper, we focus on the application of OtA computation in FL and considerably extend the
system to a more general multi-task OtA FL scenario.

The rest of this paper is organized as follows. 
Section \ref{sec:system_model} introduces the multi-task FL and communication models in a Cell-free mMIMO system. 
In Section \ref{sec:model_aggregation}, we describe the proposed OtA model aggregation strategies for three cooperation levels of APs. 
We present the OtA model aggregation strategy for a Cellular mMIMO system in Section \ref{sec:Cellular}.
In Section \ref{sec:simulation}, simulation results are presented to compare Cell-free mMIMO and Cellular mMIMO in terms of  FL convergence performance.
Finally, the main conclusions are provided in Section \ref{sec:conclusion}.

Notations: In this paper, scalars, vectors, and matrices are represented by italic letters, boldface lower-case letters, and boldface uppercase letters, respectively. $\mathbf{v}^{\mathrm{T}}$, $\mathbf{v}^{\mathrm{H}}$, and $\mathbf{v}^{*}$ denote the transpose, conjugate transpose,
and conjugate of a vector $\mathbf{v}$, respectively. 
$\|\mathbf{v}\|$ denotes the $\ell_{2}$ norm of a vector $\mathbf{v}$.
$[\mathbf{v}]_{i:j}$ denotes the sub-vector of a vector $\mathbf{v}$ with entries indexed from $i$ to $j$.  
$|\mathcal{D}|$ denotes the cardinality of a set $\mathcal{D}$.
$\mathbf{I}_{N}$ represents a $N\times N$ identity matrix and  $\mathbf{1}_{N}$ represents a $N\times 1$ all-one vector.
$\mathbb{C}^{M\times N}$ and $\mathbb{R}^{M\times N}$  represent the space of a complex-valued matrix and a real-valued matrix, respectively. 
$\mathbb{E}\{\cdot\}$ is
the statistical expectation and $\mathrm{tr}(\cdot)$ is the trace operation. 
$\mathcal{N_{\mathbb{C}}}\left(\mathbf{0}, \mathbf{R}\right)$ denotes the distribution of a multivariate circularly symmetric complex Gaussian variable with zero mean and covariance matrix $\mathbf{R}$. The important notations used in this paper are listed in Table \ref{tab1}. 

\section{System Model}
\label{sec:system_model}
We consider a Cell-free mMIMO system consisting of $L$ 
multi-antenna APs, each equipped with $N$ antennas. All APs cooperate to serve the wireless FL devices via high-speed fronthaul links connected to a central processing unit (CPU). There are $G$ FL groups, where group $g$ has $K_{g}$ single-antenna FL devices. Let $\mathcal{G}$ denote the index set of FL groups.
In particular, the index set of FL devices in group $g$ is denoted by $\mathcal{K}_{g}=\{\textstyle\sum\nolimits_{i=1}^{g-1}K_{i}+1, \dots, \sum\nolimits_{i=1}^{g-1}K_{i}+K_{g}\}$; device $k$ is in group $g$ if $k \in \mathcal{K}_{g}$. Let $K$ denote the total number of FL devices in the network with $K=\textstyle\sum\nolimits_{g=1}^{G}K_g$, and
$\mathcal{K}$ the index set of all FL devices with $\mathcal{K}=\mathcal{K}_{1}\cup\mathcal{K}_{2}\cup\dots\cup\mathcal{K}_{G}$.

\begin{table}[t]
\caption{Summary of Notations}
\begin{center}
\renewcommand\arraystretch{1.3} 
\scalebox{0.98}{
\begin{tabular}{|c|c|}
\hline   
\textbf{Notation}&\textbf{Meaning}\\
\hline  
\hline
$G$ & \tabincell{c}{Number of FL groups}  \\
\hline
$K$ & \tabincell{c}{Number of FL devices in the network}  \\
\hline
$K_{g}$ & \tabincell{c}{Number of FL devices in group $g$}  \\
\hline
$L$ & \tabincell{c}{Number of APs in the Cell-free mMIMO system}  \\
\hline
$M$ & \tabincell{c}{Number of antennas per BS in the Cellular mMIMO system}  \\
\hline
$N$ & \tabincell{c}{Number of antennas per AP}  \\
\hline
$P_k$ & \tabincell{c}{Maximum transmit power of device $k$} \\
\hline
$T$ &\tabincell{c}{Number of training rounds} \\
\hline 
$\widetilde{\mathcal{D}}_{g}$ & \tabincell{c}{Global dataset in group $g$} \\
\hline
$\mathcal{D}_{k}$ & \tabincell{c}{Local dataset of device $k$} \\
\hline
$\mathcal{G}$ & Index set of FL groups \\
\hline
$\mathcal{K}$ & \tabincell{c}{Index set of all FL devices} \\
\hline
$\mathcal{K}_{g}$ & \tabincell{c}{Index set of FL devices in group $g$} \\
\hline
$\delta^2$ &  Noise power \\
\hline
$\eta_{g}$ & Learning rate of the FL task in group $g$ \\
\hline
$\textstyle\gamma_{kg}$ & \tabincell{c}{$|\mathcal{D}_{k}|/|\widetilde{\mathcal{D}}_{g}|$} \\
\hline
$\omega_{g}$ & Weighting factor that represents the priority of group $g$ \\
\hline
$\tau_p$ & \tabincell{c}{Number of time-frequency samples for channel estimation \\ in a coherence block}\\
\hline
$\tau_u$ & \tabincell{c}{Number of time-frequency samples for  uplink  data \\transmission in a coherence block} \\
\hline
$\mathbf{C}_{kl}$ & \tabincell{c}{Spatial correlation matrix of $\widetilde{\mathbf{h}}_{kl}$} \\
\hline
$\mathbf{R}_{kl}$ & Spatial correlation matrix of $\mathbf{h}_{kl}$ \\
\hline
$\mathbf{h}_{kl}$ & Wireless channel between AP $l$ and device $k$ \\
\hline
$\widehat{\mathbf{h}}_{kl}$  &  \tabincell{c}{MMSE estimate of $\mathbf{h}_{kl}$}  \\
\hline
 $\widetilde{\mathbf{h}}_{kl}$ & \tabincell{c}{Estimate error of $\mathbf{h}_{kl}$} \\
\hline
 $\widetilde{\bm{\theta}}_{g}^{t}$ & \tabincell{c}{Desired global model parameter vector of group $g$ \\ at training round $t$} \\
\hline
$\textstyle\widehat{\bm{\theta}}_{g}^{t}$ & \tabincell{c}{Recovered global model parameter vector of group $g$ \\ at training round $t$} \\
\hline
 $\bm{\theta}_{k}^{t}$ & \tabincell{c}{Local model parameter vector of device $k$ \\ at training round $t$} \\
\hline
\end{tabular}}
\label{tab1}
\end{center}
\end{table}

\subsection{FL Model}
We assume that different groups have different FL tasks.
 To be specific, device $k$ has a local dataset $\mathcal{D}_{k}$ with cardinality $|\mathcal{D}_{k}|$. If device $k$ is in group $g$, it has the local loss function
\begin{align}
F_{k}\left(\widetilde{\bm{\theta}}_{g};\mathcal{D}_{k}\right)=\frac{1}{|\mathcal{D}_{k}|}\sum_{\mathbf{x} \in \mathcal{D}_{k}} f_{g}\left(\widetilde{\bm{\theta}}_{g};\mathbf{x}\right),
\end{align}
where $\mathbf{x}$ denotes a labeled data sample, $\widetilde{\bm{\theta}}_{g}\in \mathbb{R}^{D}$ is the global model parameter vector of group $g$, and $f_{g}\left(\widetilde{\bm{\theta}}_{g};\mathbf{x}\right)$ is the sample-wise loss function with regard to~$\mathbf{x}$.
Let $\widetilde{\mathcal{D}}_{g}$ denote the global dataset in group $g$ with cardinality $\textstyle|\widetilde{\mathcal{D}}_{g}|=\sum_{k\in \mathcal{K}_{g}}|\mathcal{D}_{k}|$.
Accordingly, the global loss function of the FL task in group $g$ is given by
\begin{align}
\label{eq:Fg}
\widetilde{F}_{g}(\widetilde{\bm{\theta}}_{g};\widetilde{\mathcal{D}}_{g})=\frac{1}{|\widetilde{\mathcal{D}}_{g}|}\sum_{k \in \mathcal{K}_{g}} |\mathcal{D}_{k}|F_{k}(\widetilde{\bm{\theta}}_{g};\mathcal{D}_{k}).
\end{align}
The learning objective of group $g$ is to obtain a global model with the parameter vector 
\begin{align}
\widetilde{\bm{\theta}}_{g}^{\mathrm{opt}}=\mathop{\arg \min_{\widetilde{\bm{\theta}}_{g}}}~& \widetilde{F}_{g}(\widetilde{\bm{\theta}}_{g};\widetilde{\mathcal{D}}_{g}).
\end{align}

During the entire FL training process, the FL devices do not transmit training data but only exchange model parameters with the Cell-free mMIMO system. Each group iteratively updates the model parameter vector for $T$ training rounds. At training round $t$, $1\le t \le T$, the training of FL in group $g$ consists of the following steps:
\begin{itemize}
	\item \textit{Model broadcast}: The APs broadcast the current recovered global model parameter vector $\textstyle\widehat{\bm{\theta}}_{g}^{t}$ to FL devices in group $g$ through error-free channels \cite{lin2021deploying}. Notice that $\textstyle\widehat{\bm{\theta}}_{g}^{1}=\widetilde{\bm{\theta}}_{g}^{1}$ denotes the initial global model parameter vector.
	\item \textit{Local model parameter update}: Each device updates its model parameter vector by computing its local gradient based on its local training dataset. Specifically, the local model parameter vector at device $k$, $ \forall k\in \mathcal{K}_{g}$, is computed as
 \begin{align}
 \label{eq:varphikt}
\bm{\theta}_{k}^{t}=\widehat{\bm{\theta}}_{g}^{t} - \eta_{g}\nabla F_{k}(\widehat{\bm{\theta}}_{g}^{t};\mathcal{D}_{k}),
\end{align}
where $\eta_{g}$ is the learning rate of the FL task in group $g$, and $\nabla F_{k}(\widehat{\bm{\theta}}_{g}^{t};\mathcal{D}_{k})\in \mathbb{R}^{D}$ denotes the gradient of $F_{k}\left(\widetilde{\bm{\theta}}_{g};\mathcal{D}_{k}\right)$ at $\widetilde{\bm{\theta}}_{g}=\widehat{\bm{\theta}}_{g}^{t}$.
\item \textit{Model aggregation}: The FL devices transmit their local model parameter vectors to the APs over wireless channels to update the global model. The desired global model parameter vector is the weighted sum of the local model parameter vectors, which is denoted by 
 \begin{align}
 \label{eq:rgt}
\widetilde{\bm{\theta}}_{g}^{t+1}=\sum_{k\in \mathcal{K}_{g}}\gamma_{kg}\bm{\theta}_{k}^{t},
\end{align}
where $\textstyle\gamma_{kg}=|\mathcal{D}_{k}|/|\widetilde{\mathcal{D}}_{g}|$.
The CPU aims to recover the desired global model parameter vector from the received signals at the APs.
However, due to channel fading and noise, there will be distortions between $\widetilde{\bm{\theta}}_{g}^{t+1}$ and the recovered global model parameter vector $\widehat{\bm{\theta}}_{g}^{t+1}$. 
\end{itemize}

\subsection{Channel Model}
 The wireless channel between AP $l$ and device $k$ is denoted by $\mathbf{h}_{kl}\in \mathbb{C}^{N}$. Under the assumption of Rayleigh fading channels,
we have 
\begin{align}
\mathbf{h}_{kl}\sim \mathcal{N}_{\mathbb{C}}\left(\mathbf{0}, \mathbf{R}_{kl}\right),
\end{align}
where $\mathbf{R}_{kl}\in \mathbb{C}^{N\times N}$ is the spatial correlation matrix and ${\beta_{kl}=\mathrm{tr}\left(\mathbf{R}_{kl}\right)/N}$ is the large-scale fading coefficient that describes geometric pathloss and shadowing. 

\subsection{Channel Estimation}
We adopt the block fading channel model in which
time-frequency resources are divided into multiple coherence blocks. 
The channels remain static and frequency-flat within a coherence block but may change over different coherence blocks.
The time-division duplex (TDD) protocol is adopted for channel estimation and data transmission.
In each coherence block, $\tau_{p}$ time-frequency samples are used for channel estimation and  $\tau_{u}$ time-frequency samples are used for uplink signal transmission.\footnote{Coherence blocks are called resource blocks in 4G/5G and are the same size for all users, even if some have channels supporting larger blocks.
The block size is standardized to define the minimum coherence time and coherence bandwidth for any user capable of connecting to the network~\cite{bjornson2017massive}.}
Specifically, each wireless device is arbitrarily assigned a pilot signal selected from $\tau_{p}$ mutually orthogonal pilot signals. 
Let $\bm{\phi}_{k}\in \mathbb{C}^{\tau_p}$ denote the pilot signal used by device $k$, and $\mathcal{P}_{k}$ denote the set of FL devices that use the same pilot signal as device $k$.
We have $\bm{\phi}_{k}^{\mathrm{H}}\bm{\phi}_{i}=0$,  $\forall i\notin \mathcal{P}_{k}$, and $\|\bm{\phi}_{k}\|^{2}=\tau_p$. 
During the channel estimation phase, the received signal at AP $l$ is given by
\begin{align}
\mathbf{Y}_{l}^{\mathrm{pilot}}=\sum_{i\in \mathcal{K}}\sqrt{p_{i}}\mathbf{h}_{il}\bm{\phi}_{i}^{\mathrm{H}} + \mathbf{N}_{l},
\end{align}
where $p_{i}$ is the pilot transmit power of device $i$ and $\mathbf{N}_{l}~\in~\mathbb{C}^{N\times \tau_p}$ is the noise matrix with i.i.d. $\mathcal{N}_{\mathbb{C}}\left(0, \delta^2\right)$ entries.
Then AP $l$ correlates the received signal with the normalized pilot signal of device $k$,
$\bm{\phi}_{k}/\sqrt{\tau_{p}}$, and obtains
\begin{align}
\mathbf{y}_{kl}^{\mathrm{pilot}}&=\sum_{i\in \mathcal{K}}\sqrt{\frac{{p_{i}}}{{\tau_{p}}}}\mathbf{h}_{il}\bm{\phi}_{i}^{\mathrm{H}}\bm{\phi}_{k} + \frac{1}{\sqrt{\tau_{p}}}\mathbf{N}_{l}\bm{\phi}_{k} \nonumber \\
&= \sum_{i\in\mathcal{P}_{k}}\sqrt{p_{i}\tau_{p}}\mathbf{h}_{il} + \mathbf{n}_{kl},
\end{align}
where $\textstyle\mathbf{n}_{kl}= \frac{1}{\sqrt{\tau_{p}}}\mathbf{N}_{l}\bm{\phi}_{k}\sim \mathcal{N}_{\mathbb{C}}\left(\mathbf{0}, \delta^2\mathbf{I}_N\right)$.
The MMSE estimate of $\mathbf{h}_{kl}$ is given by
\begin{align}
\widehat{\mathbf{h}}_{kl}=\sqrt{p_{k}\tau_{p}}\mathbf{R}_{kl}\bm{\Xi}_{kl}^{-1}\mathbf{y}_{kl}^{\mathrm{pilot}},
\end{align}
where $\textstyle\bm{\Xi}_{kl}=\mathbb{E}\{\mathbf{y}_{kl}^{\mathrm{pilot}}(\mathbf{y}_{kl}^{\mathrm{pilot}})^{\mathrm{H}}\}=\sum_{i\in\mathcal{P}_{k}}p_{i}\tau_{p}\mathbf{R}_{il} + \delta^2\mathbf{I}_{N}$. 
The MMSE estimate $\widehat{\mathbf{h}}_{kl}$ is distributed as  $\widehat{\mathbf{h}}_{kl}\sim \mathcal{N}_{\mathbb{C}}\left(\mathbf{0}, \mathbf{B}_{kl}\right)$, where
\begin{align}
\mathbf{B}_{kl}=\mathbb{E}\left\{\widehat{\mathbf{h}}_{kl}\widehat{\mathbf{h}}_{kl}^{\mathrm{H}}\right\}=
{p_{k}\tau_{p}}\mathbf{R}_{kl}\bm{\Xi}_{kl}^{-1}\mathbf{R}_{kl}.
\end{align}
The estimate error $\widetilde{\mathbf{h}}_{kl}= \mathbf{h}_{kl} - \widehat{\mathbf{h}}_{kl}$ is independent of $\widehat{\mathbf{h}}_{kl}$  and is distributed as $\widetilde{\mathbf{h}}_{kl}\sim \mathcal{N}_{\mathbb{C}}\left(\mathbf{0}, \mathbf{C}_{kl}\right)$, where $\mathbf{C}_{kl} = \mathbf{R}_{kl}-\mathbf{B}_{kl}$.

\section{OtA  Model Aggregation for Three Levels of Receiver Cooperation}
\label{sec:model_aggregation}
We adopt OtA model aggregation for communication-efficient FL model aggregation. At training round $t$, 
before uplink transmission, the local model parameter vector $\bm{\theta}_{k}^{t}$ is first normalized at device $k$ as 
\begin{align}
\label{normalization}
\mathbf{s}_{k}^{t}=\frac{\bm{\theta}_{k}^{t}- \bar{\theta}_{k}^{t}\mathbf{1}_{D}}{\nu_{k}^{t}},
\end{align}
where $\bar{\theta}_{k}^{t}$ and $\nu_{k}^{t}$ are the mean and standard deviation of the $D$ entries of $\bm{\theta}_{k}^{t}$, respectively. 
The mean and variance are computed as
\begin{align}
\bar{\theta}_{k}^{t}&=\frac{1}{D}\sum_{d=1}^{D}\theta_{k,d}^{t},\\
\left(\nu_{k}^{t}\right)^{2}&=\frac{1}{D}\sum_{d=1}^{D}\left(\theta_{k,d}^{t}-\bar{\theta}_{k}^{t}\right)^{2},
\end{align}
where $\theta_{k,d}^{t}$, $1\le d\le D$, is the $d^{\mathrm{th}}$ entry of $\bm{\theta}_{k}^{t}$. 
To mitigate the fronthaul overhead, at each training round, local mean and standard deviation statistics are assumed to be transmitted from the FL devices to the CPU  directly through error-free channels.
The normalized local gradient vector $\mathbf{s}_{k}^{t}$ is converted to $D$-slot transmit signals $\{s_{k,d}^{t}: d=1,\dots,D\}$,  where $s_{k,d}^{t}$ is the $d^{\mathrm{th}}$ entry of $\mathbf{s}_{k}^{t}$, which are then transmitted to the APs slot-by-slot. In particular, we have $\mathbb{E}\{s_{k,d}^{t}\}=0$ and  $\mathbb{E}\{|s_{k,d}^{t}|^2\}=1$.
 
During OtA model aggregation, all FL devices upload  normalized
local model parameters simultaneously.
At time slot $d$ of training round $t$, the received signal at AP $l$ is given by
\begin{align}
\mathbf{y}_{l,d}^{t}=\sum_{k\in \mathcal{K}}
\mathbf{h}_{kl,d}^{t}b_{k,d}^{t}s_{k,d}^{t} + \mathbf{n}_{l,d}^{t},
\end{align}
where $b_{k,d}^t$ is the transmit coefficient at device $k$, and ${\mathbf{n}_{l,d}^{t}\sim \mathcal{N}_{\mathbb{C}}\left(\mathbf{0}, \delta^2\mathbf{I}_N\right)}$ is the noise vector at AP $l$.
Due to the normalization step in (\ref{normalization}), we have 
${\mathbb{E}\{\left|b_{k,d}^{t}s_{k,d}^{t}\right|^2\}=|b_{k,d}^{t}|^2\le P_k}$, where $P_k$ is the maximum transmit power of device $k$.  This facilitates the design of transmit coefficients.

At training round $t$, the APs can send $\{\mathbf{y}_{l,d}^{t}: l=1,\dots,L, d=1,\dots,D\}$  to the CPU via fronthaul links for centrally recovering $\{\tilde{\theta}_{g,d}^{t+1}:g=1,\dots,G, d=1,\dots,D\}$, where $\tilde{\theta}_{g,d}^{t+1}$ is the $d^{\mathrm{th}}$ entry of the desired global model parameter vector $\widetilde{\bm{\theta}}_{g}^{t+1}$, or 
 locally recover $\{\tilde{\theta}_{g,d}^{t+1}\}$ and share these local estimates with the CPU. 
  In this paper, we consider three levels of AP cooperation and develop optimal OtA  model  aggregation strategies for them respectively.

To facilitate the design of OtA model  aggregation, we first analyze the convergence of FL in the presence of OtA  model aggregation errors. For brevity, we denote $\widetilde{F}_{g}(\widetilde{\bm{\theta}}_{g};\widetilde{\mathcal{D}}_{g})$ as $\widetilde{F}_{g}(\widetilde{\bm{\theta}}_{g})$, $\forall g \in \mathcal{G}$, hereinafter.
Following~\cite{friedlander2012hybrid, bottou2018optimization}, we make the following assumptions on the loss functions.

\textit{Assumption 1: The global loss function $\widetilde{F}_{g}(\widetilde{\bm{\theta}}_{g})$, $\forall g\in \mathcal{G}$, is continuously differentiable, and its gradient $\nabla\widetilde{F}_{g}(\widetilde{\bm{\theta}}_{g})$ is Lipschitz continuous with constant $\chi_{g}>0$, such that for any $\widetilde{\bm{\theta}}_{g}, \widetilde{\bm{\theta}}^{\prime}_{g}\in \mathbb{R}^{D}$, we have
\begin{align}
    \left\| \nabla\widetilde{F}_{g}(\widetilde{\bm{\theta}}_{g})-\nabla\widetilde{F}_{g}(\widetilde{\bm{\theta}}_{g}^{\prime})\right\| \le \chi_{g} \left\|\widetilde{\bm{\theta}}_{g} -  \widetilde{\bm{\theta}}^{\prime}_{g}\right\|,
\end{align}
which can be equally expressed as
\begin{align}
\label{eq:Lipschitz}
   &\widetilde{F}_{g}(\widetilde{\bm{\theta}}_{g})- \widetilde{F}_{g}(\widetilde{\bm{\theta}}^{\prime}_{g})\le   \nonumber \\ 
   &\quad \quad \quad  \quad \quad \left(\widetilde{\bm{\theta}}_{g} -  \widetilde{\bm{\theta}}^{\prime}_{g}\right)^{\mathrm{T}}\nabla\widetilde{F}_{g}(\widetilde{\bm{\theta}}^{\prime}_{g}) + \frac{\chi_{g}}{2}\left\|\widetilde{\bm{\theta}}_{g} -  \widetilde{\bm{\theta}}^{\prime}_{g}\right\|^{2}.
\end{align}
}

\textit{Assumption 2: The global loss function $\widetilde{F}_{g}(\widetilde{\bm{\theta}}_{g})$, $\forall g\in \mathcal{G}$, is strongly convex with parameter  $\xi_{g}>0$, such that for  any $\widetilde{\bm{\theta}}_{g}, \widetilde{\bm{\theta}}^{\prime}_{g}\in \mathbb{R}^{D}$, we have
\begin{align}
   &\widetilde{F}_{g}(\widetilde{\bm{\theta}}_{g})- \widetilde{F}_{g}(\widetilde{\bm{\theta}}^{\prime}_{g})\ge   \nonumber \\ 
   &\quad \quad \quad  \quad \quad \left(\widetilde{\bm{\theta}}_{g} -  \widetilde{\bm{\theta}}^{\prime}_{g}\right)^{\mathrm{T}}\nabla\widetilde{F}_{g}(\widetilde{\bm{\theta}}^{\prime}_{g}) + \frac{\xi_{g}}{2}\left\|\widetilde{\bm{\theta}}_{g} -  \widetilde{\bm{\theta}}^{\prime}_{g}\right\|^{2}.
\end{align}
}

Based on the above assumptions, we derive an upper bound on the expected difference between the training loss after $T$ training rounds, $\widetilde{F}_{g}(\widetilde{\bm{\theta}}^{T+1}_{g})$, and the optimal training loss, $\widetilde{F}_{g}(\widetilde{\bm{\theta}}^{\mathrm{opt}}_{g})$, in the following theorem.

\begin{theorem}
\label{theorem:FLconverge}
With Assumptions 1-2 and $\textstyle\eta_{g}=1/\chi_{g}$, after $T$ training rounds, we have
\begin{align}
\label{eq:T}
    &\mathbb{E}\left\{\widetilde{F}_{g}(\widehat{\bm{\theta}}^{T+1}_{g})-\widetilde{F}_{g}(\widetilde{\bm{\theta}}^{\mathrm{opt}}_{g})\right\}\le \nonumber \\
    &\ \underbrace{\lambda_{g}^{T}\mathbb{E}\left\{\widetilde{F}_{g}(\widehat{\bm{\theta}}^{1}_{g})-\widetilde{F}_{g}(\widetilde{\bm{\theta}}^{\mathrm{opt}}_{g})\right\}}_{\text{Initial gap}} + \underbrace{\sum_{t=1}^{T}\frac{\chi_{g}\lambda_{g}^{T-t}}{2}\mathbb{E}\left\{\left\|\mathbf{e}_{g}^{t+1}\right\|^{2}\right\}}_{\text{Aggregation error induced gap}},
\end{align}
where $\textstyle\lambda_{g}=1-\xi_{g}/\chi_{g}$ and $\mathbf{e}_{g}^{t+1}=\widetilde{\bm{\theta}}^{t+1}_{g} -  \widehat{\bm{\theta}}^{t+1}_{g}$.
\end{theorem}
\begin{IEEEproof}
	See Appendix \ref{Appendix1}.
\end{IEEEproof}

From (\ref{eq:T}), we see that the gap between 
$\mathbb{E}\{\widetilde{F}_{g}(\widehat{\bm{\theta}}^{T+1}_{g})\}$ and 
$\mathbb{E}\{\widetilde{F}_{g}(\widehat{\bm{\theta}}^{\mathrm{opt}}_{g})\}$ comes from two parts. The first part is due to the difference between the initial model parameter vector and the optimal model parameter vector, which vanishes when ${T\to \infty}$. The second part is caused by the model aggregation error. The expected model aggregation error can be derived as
\begin{align}
    \mathbb{E}\left\{\left\|\mathbf{e}_{g}^{t+1}\right\|^{2}\right\}=\sum_{d=1}^{D}\mathbb{E}\left\{\left(\tilde{\theta}_{g,d}^{t+1}-\hat{\theta}_{g,d}^{t+1}\right)^{2}\right\}.
\end{align}
At time slot $d$ of training round $t$, we aim to minimize $\mathbb{E}\{(\tilde{\theta}_{g,d}^{t+1}-\hat{\theta}_{g,d}^{t+1})^{2}\}$ by jointly optimizing the receive combining and transmit coefficients for different AP cooperation levels.

 \subsection{Fully Centralized Processing (Level~3)}
 \label{sec:level3}
At the highest level of AP cooperation, the CPU centrally performs channel estimation and recovers the
desired global model parameter vectors. 
After stacking the received data signals at time slot $d$ of training round $t$, the CPU obtains
\begin{align}
\mathbf{y}_{d}^{t}=\sum_{k\in \mathcal{K}}
\mathbf{h}_{k,d}^{t}b_{k,d}^{t}s_{k,d}^{t} + \mathbf{n}_{d}^{t},
\end{align}
where $\mathbf{y}_{d}^{t}=[(\mathbf{y}_{1,d}^{t})^{\mathrm{T}},\dots,(\mathbf{y}_{L,d}^{t})^{\mathrm{T}}]^{\mathrm{T}}\in\mathbb{C}^{LN}$, $\mathbf{h}_{k,d}^{t}=[(\mathbf{h}_{k1,d}^{t})^{\mathrm{T}},\dots,(\mathbf{h}_{kL,d}^{t})^{\mathrm{T}}]^{\mathrm{T}}\in\mathbb{C}^{LN}$, and $\mathbf{n}_{d}^{t} = [(\mathbf{n}_{1,d}^{t})^{\mathrm{T}},\dots,(\mathbf{n}_{L,d}^{t})^{\mathrm{T}}]^{\mathrm{T}}\in\mathbb{C}^{LN}$. After channel estimation, the CPU forms the stacked channel estimate $\widehat{\mathbf{h}}_{k,d}^{t}=[(\widehat{\mathbf{h}}_{k1,d}^{t})^{\mathrm{T}},\dots,(\widehat{\mathbf{h}}_{kL,d}^{t})^{\mathrm{T}}]^{\mathrm{T}}\in\mathbb{C}^{LN}$ with estimate error $\widetilde{\mathbf{h}}_{k,d}^{t}\sim \mathcal{N}_{\mathbb{C}}(\mathbf{0}, \mathbf{C}_{k})$, where $\mathbf{C}_{k}=\mathrm{diag}(\mathbf{C}_{k1},\dots,\mathbf{C}_{kL})$. Note that the channel statistics are assumed to be unchanged during the $T$ training rounds.

At Level~3, the combining vector $\mathbf{v}_{g,d}^{t}\in\mathbb{C}^{LN}$ is designed to recover $\tilde{\theta}_{g,d}^{t+1}$ as 
\begin{align}
\hat{\theta}_{g,d,(3)}^{t+1} &=
(\mathbf{v}_{g,d}^{t})^{\mathrm{H}}\mathbf{y}_{d}^{t} + \sum_{j\in \mathcal{K}_{g}}\gamma_{jg}\bar{\theta}_{j}^{t} \nonumber \\
&=
\sum_{k\in \mathcal{K}}
(\mathbf{v}_{g,d}^{t})^{\mathrm{H}}\mathbf{h}_{k,d}^{t}b_{k,d}^{t}s_{k,d}^{t} + {(\mathbf{v}_{g,d}^{t})^{\mathrm{H}}}\mathbf{n}_{d}^{t} 
+\! \sum_{j\in \mathcal{K}_{g}}\gamma_{jg}\bar{\theta}_{j}^{t} \nonumber \\
&=
\sum_{j\in \mathcal{K}_{g}}\left(
(\mathbf{v}_{g,d}^{t})^{\mathrm{H}}\mathbf{h}_{j,d}^{t}b_{j,d}^{t}s_{j,d}^{t} +\gamma_{jg}\bar{\theta}_{j}^{t}\right) \nonumber \\
&\quad \ +
\sum_{i\in \mathcal{K}\backslash \mathcal{K}_{g}}
{(\mathbf{v}_{g,d}^{t})^{\mathrm{H}}}\mathbf{h}_{i,d}^{t}b_{i,d}^{t}s_{i,d}^{t} +
{(\mathbf{v}_{g,d}^{t})^{\mathrm{H}}}\mathbf{n}_{d}^{t}.
\end{align}
The distortion between $\tilde{\theta}_{g,d}^{t+1}$ and $\textstyle\hat{\theta}_{g,d,(3)}^{t+1}$  is evaluated using the following MSE:
\begin{align}
\label{MSEg3}
&\mathrm{MSE}_{g,d,(3)}^{t}\left(\{b_{k,d}^{t}: k=1,\dots,K\},\mathbf{v}_{g,d}^{t}\right) \nonumber \\
&= \mathbb{E}\left\{\left.\left|\tilde{\theta}_{g,d}^{t+1} - \hat{\theta}_{g,d,(3)}^{t+1}\right|^2  \right| \left\{\widehat{\mathbf{h}}_{k,d}^{t}: k=1,\dots,K\right\}\right\} \nonumber \\
&\overset{(a)}{=}
\mathbb{E}\Bigg\{\Bigg|\sum_{j\in \mathcal{K}_{g}}
\left((\mathbf{v}_{g,d}^{t})^{\mathrm{H}}\mathbf{h}_{j,d}^{t}b_{j,d}^{t}-\gamma_{jg}\nu_{j}^{t}\right)s_{j,d}^{t} + \nonumber \\
& \quad \quad \
\sum_{i\in \mathcal{K}\backslash \mathcal{K}_{g}}
(\mathbf{v}_{g,d}^{t})^{\mathrm{H}}\mathbf{h}_{i,d}^{t}b_{i,d}^{t}s_{i,d}^{t} +
(\mathbf{v}_{g,d}^{t})^{\mathrm{H}}\mathbf{n}_{d}^{t}\Bigg|^2 \Bigg| \left\{\widehat{\mathbf{h}}_{k,d}^{t}\right\}\Bigg\} \nonumber \\
&=
\mathbb{E}\Bigg\{\Bigg|\sum_{j\in \mathcal{K}_{g}}
\left((\mathbf{v}_{g,d}^{t})^{\mathrm{H}}\left(\widehat{\mathbf{h}}_{j,d}^{t}+\widetilde{\mathbf{h}}_{j,d}^{t}\right)b_{j,d}^{t}-\gamma_{jg}\nu_{j}^{t}\right)s_{j,d}^{t} 
\nonumber \\
&  
\quad \quad \quad \quad \quad \quad \quad +\sum_{i\in \mathcal{K}\backslash \mathcal{K}_{g}}
(\mathbf{v}_{g,d}^{t})^{\mathrm{H}}\left(\widehat{\mathbf{h}}_{i,d}^{t}+\widetilde{\mathbf{h}}_{i,d}^{t}\right)b_{i,d}^{t}s_{i,d}^{t}  \nonumber \\
&\quad \quad \quad \quad \quad \quad \quad \quad \quad \quad \quad \quad \quad  +
{(\mathbf{v}_{g,d}^{t})^{\mathrm{H}}}\mathbf{n}_{d}^{t}\Bigg|^2 \Bigg| \left\{\widehat{\mathbf{h}}_{k,d}^{t}\right\}\!\Bigg\} \nonumber \\
&=\!
\sum_{j\in \mathcal{K}_{g}}\!\left(
\left|(\mathbf{v}_{g,d}^{t})^{\mathrm{H}}\widehat{\mathbf{h}}_{j,d}^{t}b_{j,d}^{t}\!-\!\gamma_{jg}\nu_{j}^{t}\right|^{2} \!+\! |b_{j,d}^{t}|^{2}(\mathbf{v}_{g,d}^{t})^{\mathrm{H}}\mathbf{C}_{j}\mathbf{v}_{g,d}^{t} \right)  \nonumber \\
&
\quad  +\sum_{i\in \mathcal{K}\backslash \mathcal{K}_{g}}\left(
\left|(\mathbf{v}_{g,d}^{t})^{\mathrm{H}}\widehat{\mathbf{h}}_{i,d}^{t}b_{i,d}^{t}\right|^{2} + |b_{i,d}^{t}|^{2}(\mathbf{v}_{g,d}^{t})^{\mathrm{H}}\mathbf{C}_{i}\mathbf{v}_{g,d}^{t} \right)   \nonumber \\
& \quad \quad \quad \quad \quad \quad  \quad \quad \quad \quad  \quad \quad \quad \ \quad \quad \quad + \delta^{2}\left\|\mathbf{v}_{g,d}^{t}\right\|^{2},
\end{align}
where (a) is because 
\begin{align}
    \tilde{\theta}_{g,d}^{t+1}=\sum_{j\in \mathcal{K}_{g}}\gamma_{jg}{\theta}_{j,d}^{t}=\sum_{j\in \mathcal{K}_{g}}\gamma_{jg}\left(s_{j,d}^{t}\nu_{j}^{t} + \bar{\theta}_{j}^{t} \right).
\end{align}

In this paper, we are interested in the weighted sum-MSE minimization problem, which can be written as
\begin{align}
\mathop{\min_{\{b_{k,d}^{t}\}, \{\mathbf{v}_{g,d}^{t}: g=1,\dots,G\}}}~& \sum_{g=1}^{G}\omega_{g}\mathrm{MSE}_{g,d,(3)}^{t}\left(\{b_{k,d}^{t}\},\mathbf{v}_{g,d}^{t}\right) \label{optimization}   \\
      \textrm{s.t.}~          
      & |b_{k,d}^{t}|^2\le P_k, k=1,\dots,K, \tag{\ref{optimization}{a}}  
\end{align}
where $\omega_{g}$ is the weighting factor used to represent the priority of group $g$. The optimization problem is non-convex due to the coupling of $\{b_{k,d}^{t}\}$ and $\{\mathbf{v}_{g,d}^{t}\}$.  To this end, we propose to alternatively optimize $\{b_{k,d}^{t}\}$ and $\{\mathbf{v}_{g,d}^{t}\}$ as follows.

\subsubsection{Optimization of $\{\mathbf{v}_{g,d}^{t}\}$}
For any given $\{b_{k,d}^{t}\}$, the unconstrained optimization subproblem is formulated as
\begin{align}
\label{optimize:vgs}
\mathop{\min_{\{\mathbf{v}_{g,d}^{t}\}}}~& \sum_{g=1}^{G}\omega_{g}\mathrm{MSE}_{g,d,(3)}^{t}\left(\{b_{k,d}^{t}\},\mathbf{v}_{g,d}^{t}\right).
\end{align}
Since $\mathbf{v}_{1,d}^{t}, \dots, \mathbf{v}_{G,d}^{t}$ are decoupled, (\ref{optimize:vgs}) can be further decomposed into $G$ subproblems.
The $g^{\mathrm{th}}$ convex subproblem is given by
\begin{align}
\label{optimize:cvg}
\mathop{\min_{\mathbf{v}_{g,d}^{t}}}~& \mathrm{MSE}_{g,d,(3)}^{t}\left(\{b_{k,d}^{t}\},\mathbf{v}_{g,d}^{t}\right).
\end{align}
The optimal $\mathbf{v}_{g,d}^{t}$ is obtained by setting the first derivative of (\ref{MSEg3}) to zero as 
\begin{align}
&\sum_{j\in \mathcal{K}_{g}}\left(
|b_{j,d}^{t}|^{2}\widehat{\mathbf{h}}_{j,d}^{t}(\widehat{\mathbf{h}}_{j,d}^{t})^{\mathrm{H}}\mathbf{v}_{g,d}^{t} - \gamma_{jg}b_{j,d}^{t}\nu_{j}^{t}\widehat{\mathbf{h}}_{j,d}^{t} + |b_{j,d}^{t}|^{2}\mathbf{C}_{j}\mathbf{v}_{g,d}^{t}\right) \nonumber \\
 &\ \ \ \ \quad \quad +
 \sum_{i\in \mathcal{K}\backslash \mathcal{K}_{g}}\left(
|b_{i,d}^{t}|^{2}\widehat{\mathbf{h}}_{i,d}^{t}(\widehat{\mathbf{h}}_{i,d}^{t})^{\mathrm{H}}\mathbf{v}_{g,d}^{t}
 + |b_{i,d}^{t}|^{2}\mathbf{C}_{i}\mathbf{v}_{g,d}^{t}\right) \nonumber \\
&\quad \quad \quad \quad \quad \quad \quad \quad \quad \quad \quad \quad \quad \quad \quad \quad  +
 \delta^{2}\mathbf{v}_{g,d}^{t} = 0,
\end{align}
which yields 
\begin{align}
\label{eq:v3}
\mathbf{v}_{g,d,(3)}^{t}&=\left(\sum_{k\in \mathcal{K}}
|b_{k,d}^{t}|^{2}\left(\widehat{\mathbf{h}}_{k,d}^{t}(\widehat{\mathbf{h}}_{k,d}^{t})^{\mathrm{H}}
 + \mathbf{C}_{k}\right)
 + \delta^{2}\mathbf{I}_{LN}
\right)^{-1} \nonumber \\
&\quad \quad \quad \quad \quad \quad \quad \quad \ \ \ \ \ \times \sum_{j\in \mathcal{K}_{g}}\gamma_{jg}b_{j,d}^{t}\nu_{j}^{t}\widehat{\mathbf{h}}_{j,d}^{t}.
\end{align}

\subsubsection{Optimization of $\{b_{k,d}^{t}\}$} For any given $\{\mathbf{v}_{g,d}^{t}\}$, the subproblem of transmit coefficient optimization (TCO) is formulated as 
\begin{align}
\mathop{\min_{\{b_{k,d}^{t}\}}}~& \sum_{g=1}^{G}\omega_{g}\mathrm{MSE}_{g,d,(3)}^{t}\left(\{b_{k,d}^{t}\},\mathbf{v}_{g,d}^{t}\right)  \label{optimization:p}   \\
      \textrm{s.t.}~          
      & |b_{k,d}^{t}|^2\le P_k, k=1,2,\dots,K. \tag{\ref{optimization:p}{a}}  
\end{align}
Since $b_{1,d}^{t}, \dots, b_{K,d}^{t}$ are decoupled, (\ref{optimization:p}) can be further decomposed into $K$ subproblems. When $k\in \mathcal{K}_{g}$, the subproblem associated with device $k$ is convex and is expressed as 
\begin{align}
&\mathop{\min_{b_{k,d}^{t}}} 
\omega_{g} \left(
\left|(\mathbf{v}_{g,d}^{t})^{\mathrm{H}}\widehat{\mathbf{h}}_{k,d}^{t}b_{k,d}^{t}\!-\!\gamma_{kg}\nu_{k}^{t}\right|^{2} \!+\! |b_{k,d}^{t}|^{2}(\mathbf{v}_{g,d}^{t})^{\mathrm{H}}\mathbf{C}_{k}\mathbf{v}_{g,d}^{t} \right) \nonumber \\
&   + \sum_{q=1,q\ne g}^{G}\omega_{q} \left(
\left|(\mathbf{v}_{q,d}^{t})^{\mathrm{H}}\widehat{\mathbf{h}}_{k,d}^{t}b_{k,d}^{t}\right|^{2} + |b_{k,d}^{t}|^{2}(\mathbf{v}_{q,d}^{t})^{\mathrm{H}}\mathbf{C}_{k}\mathbf{v}_{q,d}^{t} \right)  \label{optimization:P1}   \\   
      & \quad \quad \quad \quad \quad  \quad \quad \quad \quad \textrm{s.t.}~ |b_{k,d}^{t}|^2\le P_k. \tag{\ref{optimization:P1}{a}}  
\end{align}

Associating a Lagrange multiplier $\mu_{k,d}^{t}\ge 0$ to the constraint in (\ref{optimization:P1}a), we obtain the  Lagrangian function 
\begin{align}
&\mathcal{L}(b_{k,d}^{t}) \nonumber \\
&= \omega_{g} \left(
\left|(\mathbf{v}_{g,d}^{t})^{\mathrm{H}}\widehat{\mathbf{h}}_{k,d}^{t}b_{k,d}^{t}-\gamma_{kg}\nu_{k}^{t}\right|^{2} \!+ |b_{k,d}^{t}|^{2}(\mathbf{v}_{g,d}^{t})^{\mathrm{H}}\mathbf{C}_{k}\mathbf{v}_{g,d}^{t} \right) 
\nonumber \\
& \ \   + \sum_{q=1,q\ne g}^{G}\omega_{q} \left(
\left|(\mathbf{v}_{q,d}^{t})^{\mathrm{H}}\widehat{\mathbf{h}}_{k,d}^{t}b_{k,d}^{t}\right|^{2} \!+ |b_{k,d}^{t}|^{2}(\mathbf{v}_{q,d}^{t})^{\mathrm{H}}\mathbf{C}_{k}\mathbf{v}_{q,d}^{t} \right)   \nonumber \\
& \quad \quad \quad \quad \quad \quad \quad \quad \quad \quad \quad \quad \quad \quad \quad \  + \mu_{k,d}^{t}\left(|b_{k,d}^{t}|^2 - P_k\right) \nonumber \\
&= \omega_{g}\left|(\mathbf{v}_{g,d}^{t})^{\mathrm{H}}\widehat{\mathbf{h}}_{k,d}^{t}b_{k,d}^{t} - \gamma_{kg}\nu_{k}^{t}\right|^{2}
+  \mu_{k,d}^{t}\left(|b_{k,d}^{t}|^2 - P_k\right)+ \nonumber \\
&\sum_{q=1,q\ne g}^{G}\omega_{q}\left|(\mathbf{v}_{q,d}^{t})^{\mathrm{H}}\widehat{\mathbf{h}}_{k,d}^{t}b_{k,d}^{t}\right|^{2}  +
\sum_{p=1}^{G}\omega_{p}|b_{k,d}^{t}|^{2}(\mathbf{v}_{p,d}^{t})^{\mathrm{H}}\mathbf{C}_{k}\mathbf{v}_{p,d}^{t}.
\end{align}
The KKT conditions are given by
\begin{align}
b_{k,d}^{t}\Bigg(\sum_{p=1}^{G}\omega_{p}\bigg(\Big|(\mathbf{v}_{p,d}^{t})^{\mathrm{H}}\widehat{\mathbf{h}}_{k,d}^{t}\Big|^{2} + (\mathbf{v}_{p,d}^{t})^{\mathrm{H}}\mathbf{C}_{k}&\mathbf{v}_{p,d}^{t}\bigg)  \nonumber \\
 \quad \quad \quad  + \mu_{k,d}^{t}\Bigg) 
- \omega_{g}\gamma_{kg}\nu_{k}^{t}(\widehat{\mathbf{h}}_{k,d}^{t})^{\mathrm{H}}\mathbf{v}_{g,d}^{t}&=0, \label{KKTs1}  \\
  \mu_{k,d}^{t}\left(|b_{k,d}^{t}|^2-P_k\right) &= 0, \label{KKTs2} \\
 \ \ \ |b_{k,d}^{t}|^2 &\le P_k. 
\end{align}
The first-order optimality condition in (\ref{KKTs1}) yields 
\begin{align}
\label{bkopt}
&b_{k,d,(3)}^{t} \nonumber \\
&=\frac{\omega_{g}\gamma_{kg}\nu_{k}^{t}(\widehat{\mathbf{h}}_{k,d}^{t})^{\mathrm{H}}\mathbf{v}_{g,d}^{t}}{\sum_{p=1}^{G}\omega_{p}\left(\left|(\mathbf{v}_{p,d}^{t})^{\mathrm{H}}(\widehat{\mathbf{h}}_{k,d}^{t})\right|^{2} + (\mathbf{v}_{p,d}^{t})^{\mathrm{H}}\mathbf{C}_{k}\mathbf{v}_{p,d}^{t}\right) + \mu_{k,d}^{t}}.
\end{align}
 Plugging $b_{k,d,(3)}^{t}$ into the complementary slackness condition in (\ref{KKTs2}), we have
\begin{align}
\label{mukopt}
&\mu_{k,d,(3)}^{t}=\max\Bigg(0, \frac{\omega_{g}\gamma_{kg}\nu_{k}^{t}\left|(\mathbf{v}_{g,d}^{t})^{\mathrm{H}}\widehat{\mathbf{h}}_{k,d}^{t}\right|}{\sqrt{P_k}}  \nonumber \\
&\quad  - \sum_{p=1}^{G}\omega_{p}\left(\left|(\mathbf{v}_{p,d}^{t})^{\mathrm{H}}\widehat{\mathbf{h}}_{k,d}^{t}\right|^{2} + (\mathbf{v}_{p,d}^{t})^{\mathrm{H}}\mathbf{C}_{k}\mathbf{v}_{p,d}^{t}\right)\Bigg).
\end{align}
From (\ref{bkopt}) and (\ref{mukopt}), we see that $b_{k,d,(3)}^{t}$ depends on the channel conditions and the standard deviation~$\nu_{k}^{t}$.
Recall that we assume the channels remain static within a coherence block but may vary over different coherence blocks. Hence,  the transmit coefficients only need to be optimized when the coherence block or training round changes. To mitigate
the fronthaul overhead, we assume that the CPU broadcasts $\{b_{k,d,(3)}^{t}\}$ to the FL devices directly via error-free channels.

The alternating optimization algorithm operates by iteratively updating the combining vectors and transmit coefficients. In each iteration, we update
$\{\mathbf{v}_{g,d}^{t}\}$ as $\{\mathbf{v}_{g,d,(3)}^{t}\}$  and update $\{b_{k,d}^t\}$ as $\{b_{k,d,(3)}^t\}$. The algorithm is terminated if the decrease of the weighted sum-MSE is less than a threshold $\epsilon$ or the number of iterations reaches a maximum number $I_{\mathrm{max}}$.
The convergence of the proposed alternating optimization algorithm will be numerically examined in Section~\ref{sec:simulation}.

In summary, at Level~3, the APs send the received pilot signals $\{\mathbf{Y}_{l}^{\mathrm{pilot}}:l=1,\dots,L\}$ to the CPU via fronthaul links for centralized channel estimation, which requires fronthaul signaling of $\tau_p NL$ complex scalars per coherence block. 
At time slot $d$ of training round $t$, the data signals $\left\{\mathbf{y}_{l,d}^{t}:l=1,\dots,L\right\}$ are sent from the APs to the CPU. Hence, 
  $\tau_u NL$ complex scalars for the data signals are sent via fronthaul links per coherence block. 
Moreover, $KLN^2/2$ complex scalars for the Hermitian statistical matrices  ${\{\mathbf{R}_{kl}:k=1,\dots,K, l=1,\dots,L\}}$ are needed at the CPU for channel estimation and design of combining vectors. The fronthaul signaling for OtA model aggregation is summarized in Table \ref{tab2}.

\begin{table}[!t]
\begin{center}
\caption{Number of Complex Scalars to Send via the Fronthaul Links for OtA Model Aggregation.}
\label{tab2}
\scalebox{0.98}{
\begin{tabular}{| c | c | c | c |}
\hline
  & \tabincell{c}{Pilot/data signals \\ (each coherence \\ block)}  & \tabincell{c}{Combining   vectors \\ (each coherence \\ block/training round)}  &  Channel statistics \\
\hline
\hline
Level~3 & $(\tau_{p}+ \tau_{u})NL$ & $-$  & $KLN^{2}/2$  \\
\hline
Level~2 & $\tau_p NL + \tau_u GL$ & $GNL$ & $KLN^{2}/2$ \\
\hline
Level~1 & $\tau_{u}GL$ & $-$  & $-$ \\
\hline
\end{tabular}}
\end{center}
\end{table}

\subsection{Local Processing \& Centralized Combining Design (Level~2)}
At Level 2, the CPU performs channel estimation, design of combining vectors and transmission coefficients centrally as at Level~3. The well-designed combining vectors $\{\mathbf{v}_{g,d,(3)}^{t}\}$ are sent back to the APs to obtain local estimates of  $\{\tilde{\theta}_{g,d}^{t+1}\}$. To be specific,  at time slot $d$ of training round $t$, AP $l$ uses its local combining vector
$\mathbf{v}_{gl,d,(2)}^{t}=[\mathbf{v}_{g,d,(3)}^{t}]_{N(l-1)+1:Nl}$ to obtain the local estimate $\hat{\theta}_{gl,d,(2)}^{t+1}=(\mathbf{v}_{gl,d,(2)}^{t})^{\mathrm{H}}\mathbf{y}_{l,d}^{t}$.
The APs send the local estimates  $\{\hat{\theta}_{gl,d,(2)}^{t+1}:g=1,\dots,G, l=1,\dots,L\}$ to the CPU to obtain the recovery of $\tilde{\theta}_{g,d}^{t+1}$, which is given by
\begin{align}
\hat{\theta}_{g,d,(2)}^{t+1}=\sum_{l=1}^{L}\hat{\theta}_{gl,d,(2)}^{t+1} + \sum_{j\in \mathcal{K}_{g}}\gamma_{jg}\bar{\theta}_{j}^{t}=
\hat{\theta}_{g,d,(3)}^{t+1}.
\end{align}
We see that the same global model parameters can be recovered at Level~2 as at Level~3. 
The difference is that the APs do not send the received data signals, but instead send the local  estimates to the CPU. 
At time slot $d$ of training round~$t$, the local  estimates
$\{\hat{\theta}_{gl,d,(2)}^{t+1}:g=1,\dots,G,l=1,\dots,L\}$ are sent to the CPU. 
As such, $\tau_u GL$ complex scalars for the local  estimates are sent through fronthaul links per coherence block. 
From (\ref{eq:v3}), we see that the combining vectors $\{\mathbf{v}_{g,d,(3)}^{t}\}$ need to be updated once the channel conditions or standard deviations~$\{\nu_{k}^{t}\}$ change.
Accordingly, for a given coherence bandwidth, $GNL$ complex scalars for the combining vectors are sent from the CPU to the APs when the coherence block or training round changes. The other fronthaul signaling at Level~2 is the same as at Level~3, as listed in Table \ref{tab2}.

\subsection{Local Processing \& Simple Centralized Recovery (Level~1)}
At Level~1, the APs perform channel estimation locally and send the local estimates of the global model parameters to the CPU.
 At time slot $d$ of training round $t$, the local estimate of $\tilde{\theta}_{g,d}^{t+1}$ at AP $l$ is given by
\begin{align}
\hat{\theta}_{gl,d}^{t+1}
&=(\mathbf{v}_{gl,d}^{t})^{\mathrm{H}}\mathbf{y}_{l,d}^{t} + \sum_{j\in \mathcal{K}_{g}}\gamma_{jg}\bar{\theta}_{j}^{t} \nonumber \\
&=
\sum_{j\in \mathcal{K}_{g}}
\left((\mathbf{v}_{gl,d}^{t})^{\mathrm{H}}\mathbf{h}_{jl,d}^{t}b_{j,d}^{t}s_{j,d}^{t} + \gamma_{jg}\bar{\theta}_{j}^{t}\right) \nonumber \\
&\ \ \  +
\sum_{i\in \mathcal{K}\backslash \mathcal{K}_{g}}(\mathbf{v}_{gl,d}^{t})^{\mathrm{H}}\mathbf{h}_{il,d}^{t}b_{i,d}^{t}s_{i,d}^{t}
+ (\mathbf{v}_{gl,d}^{t})^{\mathrm{H}}\mathbf{n}_{l,d}^{t},
\end{align}
where $\mathbf{v}_{gl,d}^{t}\in \mathbb{C}^{N}$ denotes the local combining vector for group $g$ at AP $l$. 
We design $\mathbf{v}_{gl,d}^{t}$ to minimize the following MSE:
\begin{align}
\label{eq:localMSE}
&\mathbb{E}\left\{\left.\left|\tilde{\theta}_{g,d}^{t+1} - \hat{\theta}_{gl,d}^{t+1}\right|^2  \right| \left\{\widehat{\mathbf{h}}_{kl,d}^{t}: k=1,\dots,K\right\}\right\} \nonumber \\
&=
\mathbb{E}\Bigg\{\Bigg|\sum_{j\in \mathcal{K}_{g}}
\left((\mathbf{v}_{gl,d}^{t})^{\mathrm{H}}\mathbf{h}_{jl,d}^{t}b_{j,d}^{t} - \gamma_{jg}\nu_{j}^{t}\right)s_{j,d}^{t} 
+  \nonumber \\
&\ \ \sum_{i\in \mathcal{K}\backslash \mathcal{K}_{g}}(\mathbf{v}_{gl,d}^{t})^{\mathrm{H}}\mathbf{h}_{il,d}^{t}b_{i,d}^{t}s_{i,d}^{t}
+ (\mathbf{v}_{gl,d}^{t})^{\mathrm{H}}\mathbf{n}_{l,d}^{t}\Bigg|^2 \Bigg|\left\{\widehat{\mathbf{h}}_{kl,d}^{t}\right\}\Bigg\} \nonumber \\
&=\!\!
\sum_{j\in \mathcal{K}_{g}}\!\left(
\left|(\mathbf{v}_{gl,d}^{t})^{\mathrm{H}}\widehat{\mathbf{h}}_{jl,d}^{t}b_{j,d}^{t}\!-\!\gamma_{jg}\nu_{j}^{t}\right|^{2} \!\!+\! |b_{j,d}^{t}|^{2}(\mathbf{v}_{gl,d}^{t})^{\mathrm{H}}\mathbf{C}_{jl}\mathbf{v}_{gl,d}^{t} \right) \nonumber \\
& \ \ +
\sum_{i\in \mathcal{K}\backslash \mathcal{K}_{g}}\left(\left|(\mathbf{v}_{gl,d}^{t})^{\mathrm{H}}\widehat{\mathbf{h}}_{il,d}^{t}b_{i,d}^{t}\right|^{2} + |b_{i,d}^{t}|^{2}(\mathbf{v}_{gl,d}^{t})^{\mathrm{H}}\mathbf{C}_{il}\mathbf{v}_{gl,d}^{t}\right)    \nonumber \\
& \quad \quad \quad \quad \quad \quad \quad \quad \quad \quad \quad \quad \quad \quad \quad \quad + \delta^{2}\left\|\mathbf{v}_{gl,d}^{t}\right\|^{2}.
\end{align}
 By equating the first derivative of (\ref{eq:localMSE}) with regard to $\mathbf{v}_{gl,d}^{t}$ to zero, the optimal $\mathbf{v}_{gl,d}^{t}$ at Level~1 is obtained as
 \begin{align}\mathbf{v}_{gl,d,(1)}^{t}&=\left(\sum_{k\in \mathcal{K}}
|b_{k,d}^{t}|^{2}\left(\widehat{\mathbf{h}}_{kl,d}^{t}(\widehat{\mathbf{h}}_{kl,d}^{t})^{\mathrm{H}}
 + \mathbf{C}_{kl}\right)
 + \delta^{2}\mathbf{I}_{N}
\right)^{-1} \nonumber \\
&\quad \quad \quad \quad \quad \quad \quad \quad \quad \ \times \sum_{j\in \mathcal{K}_{g}}\gamma_{jg}b_{j,d}^{t}\nu_{j}^{t}\widehat{\mathbf{h}}_{jl,d}^{t},
 \end{align}
which yields the local estimate of $\tilde{\theta}_{g,d}^{t+1}$ as
\begin{align}
\hat{\theta}_{gl,d,(1)}^{t+1}&=
\sum_{k\in\mathcal{K}}
(\mathbf{v}_{gl,d,(1)}^{t})^{\mathrm{H}}\mathbf{h}_{kl,d}^{t}b_{k,d}^{t}s_{k,d}^{t}  \nonumber \\
&\quad \quad \quad \quad \ \ \ \ \ + (\mathbf{v}_{gl,d,(1)}^{t})^{\mathrm{H}}\mathbf{n}_{l,d}^{t} + \sum_{j\in \mathcal{K}_{g}}\gamma_{jg}\bar{\theta}_{j}^{t}.
\end{align}
It is not feasible to perform TCO at Level 1 as the CPU  has no channel information.
As such, the transmit coefficients are simply set to ${b_{k,d,(1)}^{t}=\sqrt{P_{k}}, k=1,\dots,K}$.
The local estimates  $\{\hat{\theta}_{gl,d,(1)}^{t+1}:l=1,\dots,L\}$ are then sent to the CPU for centralized recovery of $\tilde{\theta}_{g,d}^{t+1}$. In the absence of global channel information, the CPU simply  averages the local estimates and obtains
\begin{align}
\hat{\theta}_{g,d,(1)}^{t+1}=\frac{1}{L}\sum_{l=1}^{L}\hat{\theta}_{gl,d,(1)}^{t+1}.
\end{align}

To facilitate the evaluation of the MSE between $\tilde{\theta}_{g,d}^{t+1}$ and $\hat{\theta}_{g,d,(1)}^{t+1}$, we rewrite $\hat{\theta}_{g,d,(1)}^{t+1}$ as
 \begin{align}
&\hat{\theta}_{g,d,(1)}^{t+1} \nonumber \\
&=
\sum_{l=1}^{L}\frac{1}{L}\Bigg(\sum_{j\in\mathcal{K}_{g}}
(\mathbf{v}_{gl,d,(1)}^{t})^{\mathrm{H}}\mathbf{h}_{jl,d}^{t}b_{j,d}^{t}s_{j,d}^{t} + \sum_{j\in \mathcal{K}_{g}}\gamma_{jg}\bar{\theta}_{j}^{t} +
 \nonumber \\
& 
\ \ \ \ \ \ \sum_{i\in \mathcal{K}\backslash \mathcal{K}_{g}}(\mathbf{v}_{gl,d,(1)}^{t})^{\mathrm{H}}\mathbf{h}_{il,d}^{t}b_{i,d}^{t}s_{i,d}^{t} +   (\mathbf{v}_{gl,d,(1)}^{t})^{\mathrm{H}}\mathbf{n}_{l,d}^{t} \Bigg) \nonumber \\
&=
\sum_{j\in\mathcal{K}_{g}}
\mathbf{a}^{\mathrm{H}}\mathbf{u}_{gj,d}^{t}b_{j,d}^{t}s_{i,d}^{t} 
+ \sum_{j\in \mathcal{K}_{g}}\gamma_{jg}\bar{\theta}_{j}^{t}+
\nonumber \\
&\quad \quad \quad \quad \quad \quad \quad \ \
\sum_{i\in \mathcal{K}\backslash \mathcal{K}_{g}}\mathbf{a}^{\mathrm{H}}\mathbf{u}_{gi,d}^{t}b_{i,d}^{t}s_{i,d}^{t} +
\mathbf{a}^{\mathrm{H}}\mathbf{m}_{g,d}^{t},
\end{align}
where  $\mathbf{u}_{gi,d}^{t}=[(\mathbf{v}_{g1,d,(1)}^{t})^{\mathrm{H}}\mathbf{h}_{i1,d}^{t},\dots,(\mathbf{v}_{gL,d,(1)}^{t})^{\mathrm{H}}\mathbf{h}_{iL,d}^{t}]^{\mathrm{T}}$,
$\mathbf{m}_{g,d}^{t}=[(\mathbf{v}_{g1,d,(1)}^{t})^{\mathrm{H}}\mathbf{n}_{1,d}^{t},\dots,(\mathbf{v}_{gL,d,(1)}^{t})^{\mathrm{H}}\mathbf{n}_{L,d}^{t}]^{\mathrm{T}}$, and ${\textstyle\mathbf{a}=\frac{1}{L}\mathbf{1}_{L}}$. Then the MSE between   $\tilde{\theta}_{g,d}^{t+1}$ and $\hat{\theta}_{g,d,(1)}^{t+1}$ is expressed as
\begin{align}
\label{eq:MSEg2}
&\mathrm{MSE}_{g,d,(1)}^{t} \nonumber \\
&=\mathbb{E}\left\{\left.\left|\tilde{\theta}_{g,d}^{t+1} - \hat{\theta}_{g,d,(1)}^{t+1}\right|^2  \right| \left\{\mathbf{u}_{gk,d}^{t}:k= 1,\dots,K\right\}\right\} \nonumber \\
&= 
\sum_{j\in \mathcal{K}_{g}}
\left|\mathbf{a}^{\mathrm{H}}{\mathbf{u}}_{gj,d}^{t}b_{j,d}^{t}-\gamma_{jg}\nu_{j}^{t}\right|^{2} 
+ \nonumber \\
& \quad \quad \quad \quad \quad \quad \quad \sum_{i\in \mathcal{K}\backslash \mathcal{K}_{g}}\left|\mathbf{a}^{\mathrm{H}}\mathbf{u}_{gi,d}^{t}b_{i,d}^{t}\right|^{2}+
\delta^{2}\mathbf{a}^{\mathrm{H}}\mathbf{Z}_{g,d}^{t}\mathbf{a},
\end{align}
where $\mathbf{Z}_{g,d}^{t}=\text{diag}(\|\mathbf{v}_{g1,d,(1)}^{t}\|^{2}, \dots, \|\mathbf{v}_{gL,d,(1)}^{t}\|^{2})$.

In summary, at Level 1, the APs do not send any pilot signals and channel statistics to the CPU. $\tau_u GL$ complex scalars for local estimates ${\{\hat{\theta}_{gl,d,(1)}^{t+1}:g=1,\dots,G,l=1,\dots,L\}}$
are forwarded from the APs to the CPU through fronthaul links per coherence block,  which are listed in Table \ref{tab2}.

\section{OtA Model Aggregation Design for Cellular mMIMO Systems}
\label{sec:Cellular}
One of the main contributions of this paper is to compare Cell-free mMIMO with Cellular mMIMO in terms of FL convergence performance.
In this section, we present the optimal multi-task OtA model aggregation design for conventional Cellular mMIMO systems. Since antenna arrays are co-located in Cellular mMIMO, fully centralized processing is employed.
Assume that each mMIMO BS is equipped with $M$ antennas. Let $\check{g}$ denote the index of the BS responsible for the FL task of group $g$.
At time slot $d$ of training round $t$, the received signal at BS $\check{g}$ is given by
\begin{align}
\mathbf{y}^{t,\mathrm{BS}}_{\check{g},d}=\sum_{k\in \mathcal{K}}
\mathbf{h}^{t,\mathrm{BS}}_{k\check{g},d}b_{k,d}^{t}s_{k,d}^{t} + \mathbf{n}^{t,\mathrm{BS}}_{\check{g},d},
\end{align}
where $\mathbf{h}^{t,\mathrm{BS}}_{k\check{g},d}\in \mathbb{C}^{M}$ is the wireless channel 
between BS $\check{g}$ and FL device $k$, and $\mathbf{n}^{t,\mathrm{BS}}_{\check{g},d}\sim \mathcal{N}_{\mathbb{C}}\left(\mathbf{0}, \delta^2\mathbf{I}_M\right)$ is the noise vector at BS~$\check{g}$.

The combining vector $\mathbf{w}_{g,d}^{t}\in\mathbb{C}^{M}$ is designed to recover $\tilde{\theta}_{g,d}^{t+1}$ as 
\begin{align}
\hat{\theta}_{g,d}^{t+1,\mathrm{BS}}&=
(\mathbf{w}_{g,d}^{t})^{\mathrm{H}}\mathbf{y}^{t,\mathrm{BS}}_{\check{g},d} + \sum_{j\in \mathcal{K}_{g}}\gamma_{jg}\bar{\theta}_{j}^{t} \nonumber \\
&= 
\sum_{j\in \mathcal{K}_{g}}
\left((\mathbf{w}_{g,d}^{t})^{\mathrm{H}}\mathbf{h}^{t,\mathrm{BS}}_{j\check{g},d}b_{j,d}^{t}s_{j,d}^{t} + \gamma_{jg}\bar{\theta}_{j}^{t}\right) \nonumber \\
&\ \sum_{i\in \mathcal{K}\backslash \mathcal{K}_{g}}
{(\mathbf{w}_{g,d}^{t})^{\mathrm{H}}}\mathbf{h}^{t,\mathrm{BS}}_{i\check{g},d}b_{i,d}^{t}s_{i,d}^{t} +
{(\mathbf{w}_{g,d}^{t})^{\mathrm{H}}}\mathbf{n}^{t,\mathrm{BS}}_{\check{g},d}.
\end{align}

The MSE between  $\tilde{\theta}_{g,d}^{t+1}$ and $\hat{\theta}_{g,d}^{t+1,\mathrm{BS}}$ is given by
\begin{align}
\label{cMSEg3}
&\mathrm{MSE}^{t,\mathrm{BS}}_{g,d}\left(\{b_{k,d}^{t}\},\mathbf{w}_{g,d}^{t}\right) \nonumber \\
&= \mathbb{E}\left\{\left.\left|\tilde{\theta}_{g,d}^{t+1} - \hat{\theta}_{g,d}^{t+1,\mathrm{BS}}\right|^2  \right| \left\{\widehat{\mathbf{h}}_{k\check{g}}^{t,\mathrm{BS}}:k= 1,\dots,K\right\}\right\} \nonumber \\
&=
\sum_{j\in \mathcal{K}_{g}}\bigg(
\left|(\mathbf{w}_{g,d}^{t})^{\mathrm{H}}\widehat{\mathbf{h}}^{t,\mathrm{BS}}_{j\check{g},d}b_{j,d}^{t}-\gamma_{jg}\nu_{j}^{t}\right|^{2} + \nonumber \\
&\ \ \ \ |b_{j,d}^{t}|^{2}(\mathbf{w}_{g,d}^{t})^{\mathrm{H}}\mathbf{C}_{j\check{g}}^{\mathrm{BS}}\mathbf{w}_{g,d}^{t} \bigg) \!+\!
\sum_{i\in \mathcal{K}\backslash \mathcal{K}_{g}}\bigg(
\left|(\mathbf{w}_{g,d}^{t})^{\mathrm{H}}\widehat{\mathbf{h}}^{t,\mathrm{BS}}_{i\check{g},d}b_{i,d}^{t}\right|^{2}  \nonumber \\
&\quad \quad \quad \quad +|b_{i,d}^{t}|^{2}(\mathbf{w}_{g,d}^{t})^{\mathrm{H}}\mathbf{C}_{i\check{g}}^{\mathrm{BS}}\mathbf{w}_{g,d}^{t} \bigg)
+ \delta^{2}\left\|\mathbf{w}_{g,d}^{t}\right\|^{2},
\end{align}
where $\widehat{\mathbf{h}}_{k\check{g},d}^{t,\mathrm{BS}}$ is the MMSE estimate of ${\mathbf{h}}_{k\check{g},d}^{t,\mathrm{BS}}$, and $\mathbf{C}_{k\check{g}}^{\mathrm{BS}}$ is the covariance matrix of 
the estimate error.

The weighted sum-MSE minimization problem in the Cellular mMIMIO system is formulated as
\begin{align}
&\mathop{\min_{\{b_{k,d}^{t}\}, \{\mathbf{w}_{g,d}^{t}: g=1,\dots,G\}}}\sum_{g=1}^{G}\omega_{g}\mathrm{MSE}^{t,\mathrm{BS}}_{g,d}\left(\{b_{k,d}^{t}\},\mathbf{w}_{g,d}^{t}\right) \label{coptimization}   \\
     &\quad \quad \quad \quad \quad \textrm{s.t.}~          
       |b_{k,d}^{t}|^2\le P_k, k=1,\dots,K. \tag{\ref{coptimization}{a}}  
\end{align}
We adopt alternating optimization following similar steps of solving (\ref{optimization}). For any given $\{b_{k,d}^{t}\}$, we have
\begin{align}
\mathbf{w}_{g,d}^{t}&=\left(\sum_{k\in \mathcal{K}}
|b_{k,d}^{t}|^{2}\left(\widehat{\mathbf{h}}^{t,\mathrm{BS}}_{k\check{g},d}\left(\widehat{\mathbf{h}}^{t,\mathrm{BS}}_{k\check{g},d}\right)^{\mathrm{H}}
 \!+\! \mathbf{C}^{\mathrm{BS}}_{k\check{g}}\right)
 \!+\! \delta^{2}\mathbf{I}_{M}
\right)^{-1} \nonumber \\
&\quad \quad \quad \quad \quad \quad \quad \quad \quad \quad \ \times \sum_{j\in \mathcal{K}_{g}}\gamma_{jg}b_{j,d}^{t}\nu_{j}^{t}\widehat{\mathbf{h}}^{t,\mathrm{BS}}_{j\check{g},d}.
\end{align} 
For any given $\{\mathbf{w}_{g,d}^{t}\}$, when $k\in \mathcal{K}_{g}$, we have
\begin{align}
\label{cbkopt}
&b_{k,d}^{t}  \nonumber \\
&=\frac{\omega_{g}\gamma_{kg}\nu_{k}^{t}\left(\widehat{\mathbf{h}}^{t,\mathrm{BS}}_{k\check{g},d}\right)^{\mathrm{H}}\mathbf{w}_{g,d}^{t}}{\sum_{p=1}^{G}\omega_{p}\left(\left|(\mathbf{w}_{p,d}^{t})^{\mathrm{H}}\widehat{\mathbf{h}}^{t,\mathrm{BS}}_{k\check{p}}\right|^{2} + (\mathbf{w}_{p,d}^{t})^{\mathrm{H}}\mathbf{C}^{\mathrm{BS}}_{k\check{p}}\mathbf{w}_{p,d}^{t}\right) + \mu_{k,d}^{t}},
\end{align}
 where
\begin{align}
\label{cmukopt}
&\mu_{k,d}^{t}=\max\Bigg(0, \frac{\omega_{g}\gamma_{kg}\nu_{k}^{t}\left|(\mathbf{w}_{g,d}^{t})^{\mathrm{H}}\widehat{\mathbf{h}}^{t,\mathrm{BS}}_{k\check{g},d}\right|}{\sqrt{P_k}} \nonumber \\
&\ - \sum_{p=1}^{G}\omega_{p}\left(\left|(\mathbf{w}_{p,d}^{t})^{\mathrm{H}}\widehat{\mathbf{h}}^{t,\mathrm{BS}}_{k\check{p}}\right|^{2} + (\mathbf{w}_{p,d}^{t})^{\mathrm{H}}\mathbf{C}^{\mathrm{BS}}_{k\check{p}}\mathbf{w}_{p,d}^{t}\right)\Bigg).
\end{align}
The convergence of the above alternating optimization algorithm will be numerically examined in Section~\ref{sec:simulation}.

\section{Numerical Results}
\label{sec:simulation}
In this section, we present the MSE and FL convergence performance achieved by Cell-free mMIMO systems for the proposed three different cooperation levels, as well as by Cellular mMIMO. We consider a  $500\times 500$ m simulation area and adopt a wrap around topology to simulate a large area without unnatural boundaries. The Cellular mMIMO system has 4  square cells with a total  of $4M$ antennas. 
The APs of the Cell-free  mMIMO system are also deployed on a square grid. The total number of antennas in the Cell-free  mMIMO network is set to $4M$ for fair comparison.
Each group has the same number of FL devices, i.e., $K_{g}=K/G, g=1,\dots,G$.
The pilot length is set to $\tau_p = K/G$ and the $\tau_p$ orthogonal pilot signals are arbitrarily assigned to the FL devices in each group. Therefore, there is no intra-group pilot contamination.
The uplink pilot powers are set to $p_k=20$ dBm, $k=1,\dots,K$. 
All FL devices are assumed to have the same maximum transmit power $P_{\mathrm{max}}$, i.e., $P_{k}=P_{\mathrm{max}}, k=1,\dots,K$.
We assume a carrier frequency of 2 GHz, a bandwidth of 20 MHz, and 
a noise power of $\delta^2=-96$ dBm \cite{bjornson2019making}.

We use the same propagation model for Cell-free  mMIMO and Cellular mMIMO for fair comparison.
Following the 3GPP Urban Microcell model~\cite{3gpp2010further}, the large-scale fading is given by 
\begin{align}
\beta_{kl}[\text{dB}]= \beta_{0} - 10\alpha \text{log}_{10}\left(\frac{d_{kl}}{d_0}\right) + S_{kl},
\end{align} 
where $d_{0}=1$~m is the reference distance, $\beta_{0}=-30.5$~dB is the large-scale path loss at the reference distance, $\alpha=3.67$ is the path-loss exponent, $d_{kl}$ is the distance between AP $l$ and FL device $k$, and $S_{kl}\sim \mathcal{N}(0,4^2)$ is the shadow fading. The correlation between two shadow terms associated with AP $l$ is 
$\mathbb{E}\{S_{kl}S_{il}\}=4^2 2^{-x_{ki}/9 \ \text{m}}$, where $x_{ki}$ is the distance between FL device $k$ and FL device $i$~\cite{3gpp2010further}. The 
shadow terms associated with two different APs can be considered as uncorrelated due to the large inter-AP distance. All antenna arrays are assume to be uniform linear arrays with half-wavelength spacing. The widely used Gaussian local scattering model is adopted to model spatial channel correlation with $15^{\circ}$ angular standard deviation \cite{bjornson2017massive}.

\begin{figure}[t]
    \centering
    \subfloat[Device Distribution Mode 1: FL devices in group $g$ are uniformly distributed in cell $g$.]{\includegraphics[width=2.4in]{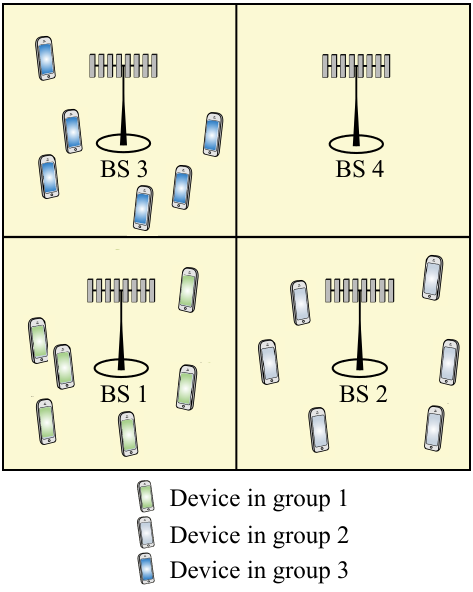}}  \\
    \subfloat[Device Distribution Mode 2: FL devices are uniformly distributed throughout the considered area.]{\includegraphics[width=2.4in]{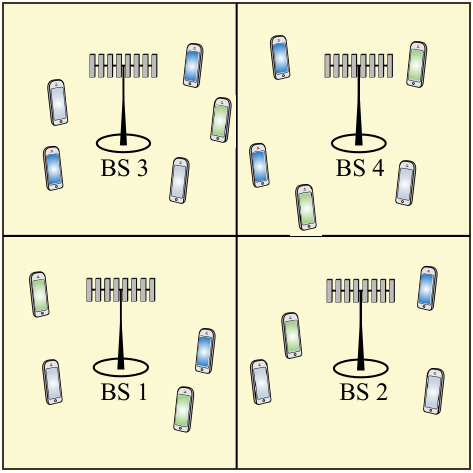}}
    \caption{Illustration of different device distribution modes with $G=3$ and $K=18$.}
    \label{fig:distribution}
\end{figure}

We consider two different device distribution modes. As shown in Fig.~\ref{fig:distribution}(a),  in Device Distribution Mode 1, the FL devices in group $g$ are uniformly distributed in cell $g$. In practice, the devices in each FL group may be distributed over multiple cells. As such, we consider Device Distribution Mode 2 as shown in Fig.~\ref{fig:distribution}(b), where the FL devices are uniformly distributed in the considered simulation area. 
Under both device distribution modes, in  Cellular mMIMO, FL devices in group $g$ are assumed to be served by BS $g$.

We assign different FL tasks to different FL groups. Group~1 has the task of fashion product recognition using the Fashion-MNIST~\cite{xiao2017fashion}  dataset;
Group~2 has the task of handwritten digit recognition using the  MNIST~\cite{lecun1998gradient} dataset; Group 3 has the task of handwritten letter recognition using the  EMNIST~\cite{cohen2017emnist} dataset.
In particular, we select data labeled with A-J from the EMNIST dataset to have 10 different labels. 
Each FL device has 500 local training samples 
which are i.i.d. drawn from the global training dataset, and trains a feedforward neural network (FNN) with an input layer of 784 nodes, a hidden layer of 60 neurons, and an output layer of 10 neurons. 
Each FNN applies a Tansig activation function to the hidden layer and a Softmax activation function to the output layer.
The loss function is cross entropy loss and the learning rate is 0.005. 
During FL training, we assume that the wireless channels remain static within a single training round, but vary across different FL training rounds.

\begin{figure}[t]
    \centering
{\includegraphics[width=3.3in]{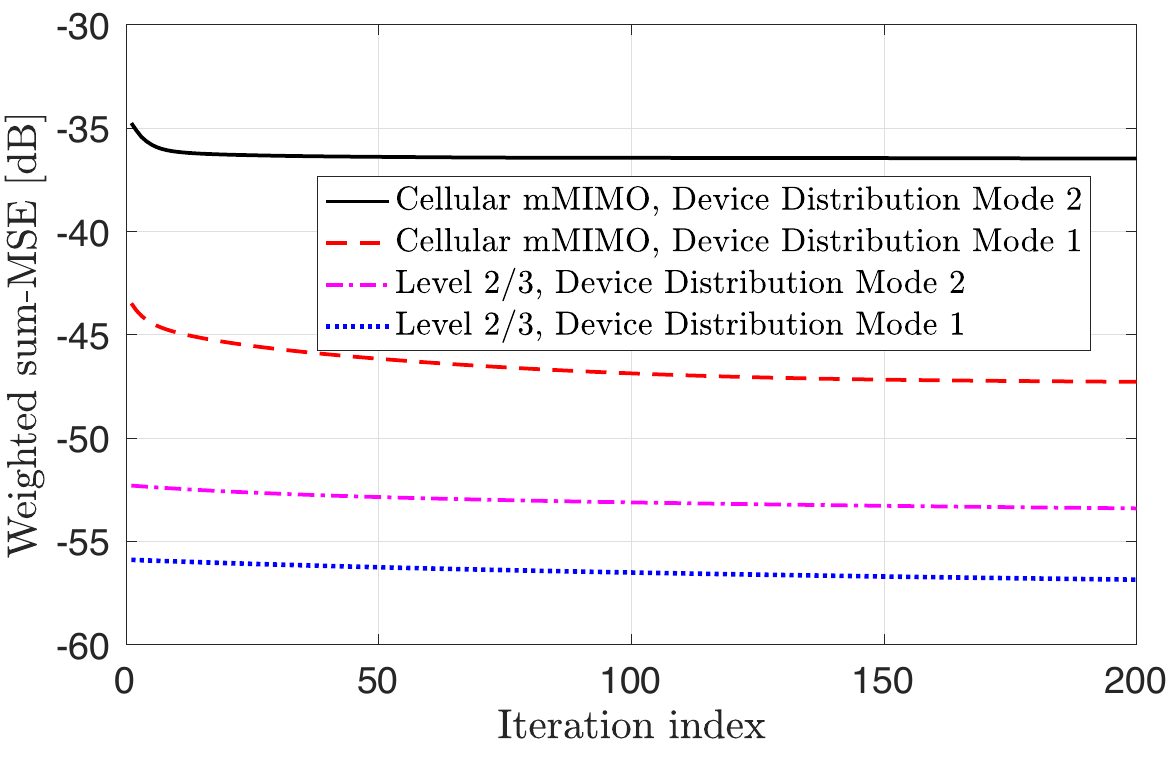}}
    \caption{Convergence performance of the proposed alternating optimization algorithms for a random channel realization with $M=64$, $L=64$, $N=4$, $K=30$, $G=3$ and ${P_{\mathrm{max}}=20}$~dBm.}
    \label{fig:convergence}
\end{figure}

\subsection{MSE Performance Evaluation}
\label{sec:MSE}
From (\ref{MSEg3}), we see that the MSE  depends on the standard deviations of the local model parameters.
Different OtA model aggregation designs may lead to different local model updates and  thus different standard deviations.
For fair comparison, we present the MSE at the first training round using the initial  global model parameters $\{\widetilde{\bm{\theta}}_{g}^{1}:g=1,\dots,G\}$. 

In Fig. \ref{fig:convergence}, we show the convergence performance of the alternating optimization algorithms proposed in Section \ref{sec:level3} and Section \ref{sec:Cellular}. The initial transmit coefficients are set to ${b_{k,d}^{t}=\sqrt{P_{k}}, k=1,\dots,K}$. 
We set the threshold of termination to  $\textstyle \epsilon=10^{-10}$ and the maximum number of iterations to $I_{\mathrm{max}}=500$. From this figure, we conclude that the proposed alternating optimization algorithms  converge after a few iterations under both device distribution modes.

\begin{figure}[t]
    \centering
    \subfloat[Device Distribution Mode 1.]{\includegraphics[width=3.3in]{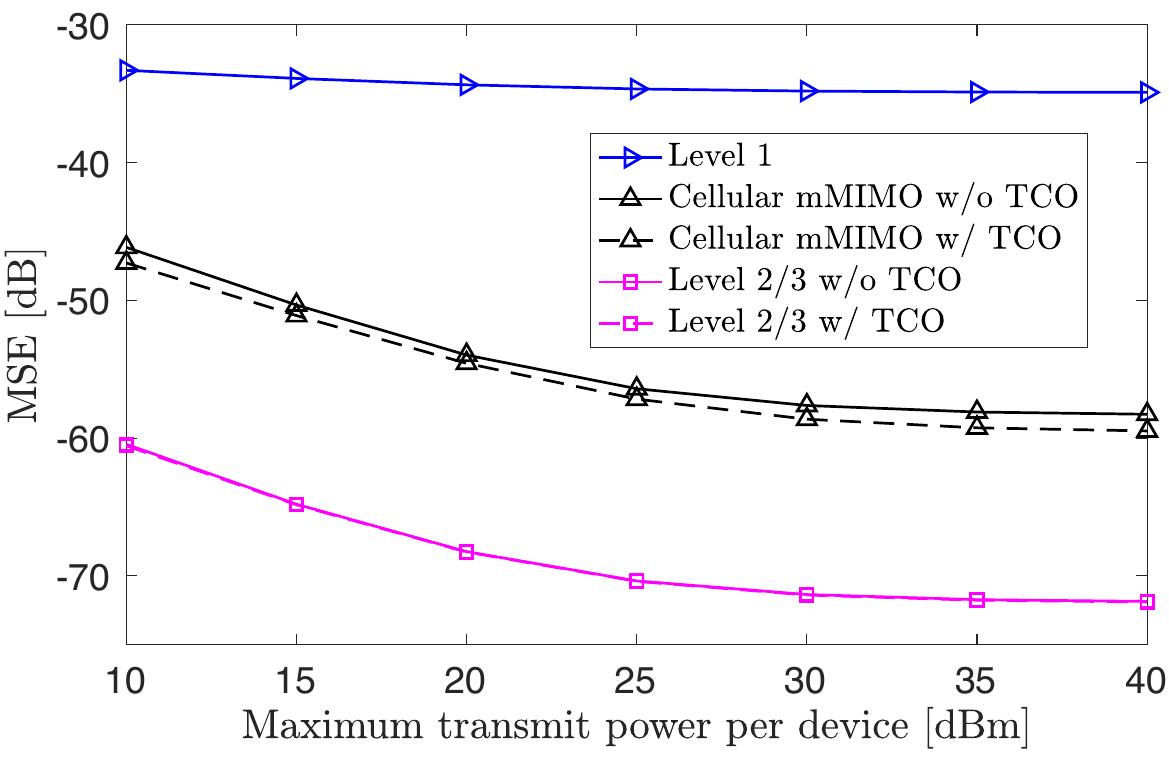}}
      \\
    \subfloat[Device Distribution Mode 2.]{\includegraphics[width=3.3in]{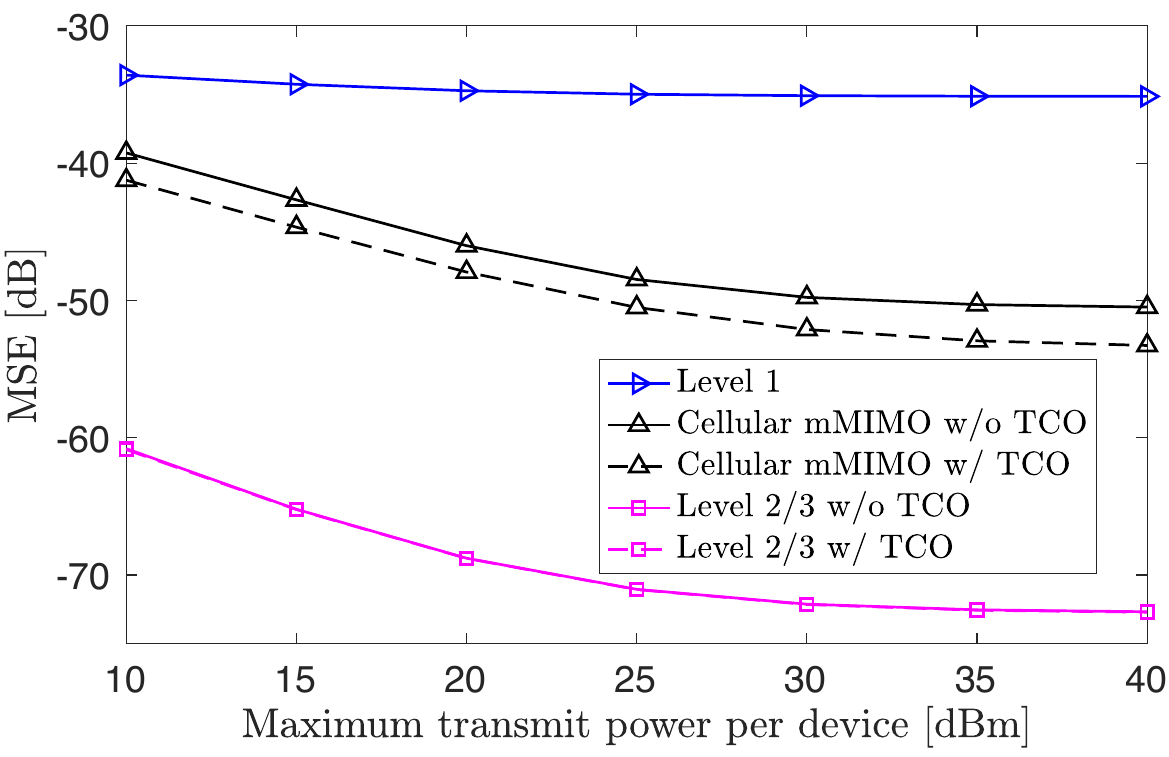}}
    \caption{MSE versus the maximum transmit power per device  with $M=64$, $L=64$, $N=4$, $K=10$ and $G=1$.}
    \label{fig:G1}
\end{figure}

In Fig.~\ref{fig:G1}, we consider one FL group and show the MSE performance of OtA model aggregation versus the maximum transmit power  per FL device. 
The results without TCO are obtained by setting the transmit coefficients to ${b_{k,d}^{t}=\sqrt{P_{k}}, k=1,\dots,K}$.
We see that the MSE of Level~2/3 is significantly lower than that of Cellular mMIMO for both device distribution modes. However, Level~1 cannot outperform Cellular mMIMO due to limited cooperation among the APs.
Moreover, we observe that the benefit of using TCO in Cell-free mMIMO is not significant for both device distribution modes due to the uniform distribution of a large number of cooperating APs. This means we don't have to rely on TCO to achieve a small MSE in the case of one FL group. 
Comparing Fig.~\ref{fig:G1}(a) and (b), we see that for Cellular mMIMO, the MSE   under Device Distribution Mode 2 is much higher than that under Device Distribution Mode 1.  On the other hand, Cell-free mMIMO achieves similar MSE under different device distribution modes.
Another important observation is that for both Cell-free mMIMO and Cellular mMIMO systems, the MSE first decreases with the maximum transmit power per wireless device and then reaches a non-zero minimum value, even if there is no inter-group interference. This is due to the presence of channel estimation errors and has been analytically proven in our previous work \cite{chen2024over}. 

\begin{figure}[!t]
    \centering
    \subfloat[Device Distribution Mode 1.]{\includegraphics[width=3.3in]{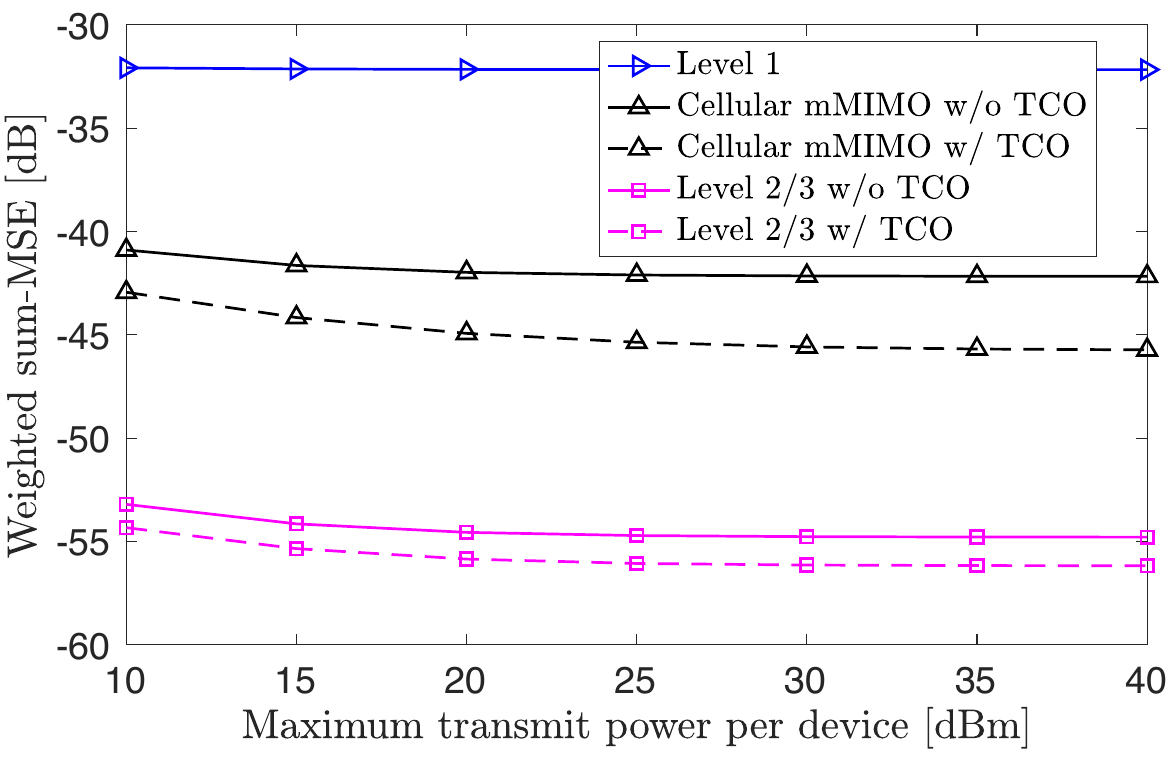}}
      \\
      \subfloat[Device Distribution Mode 2.]{\includegraphics[width=3.3in]{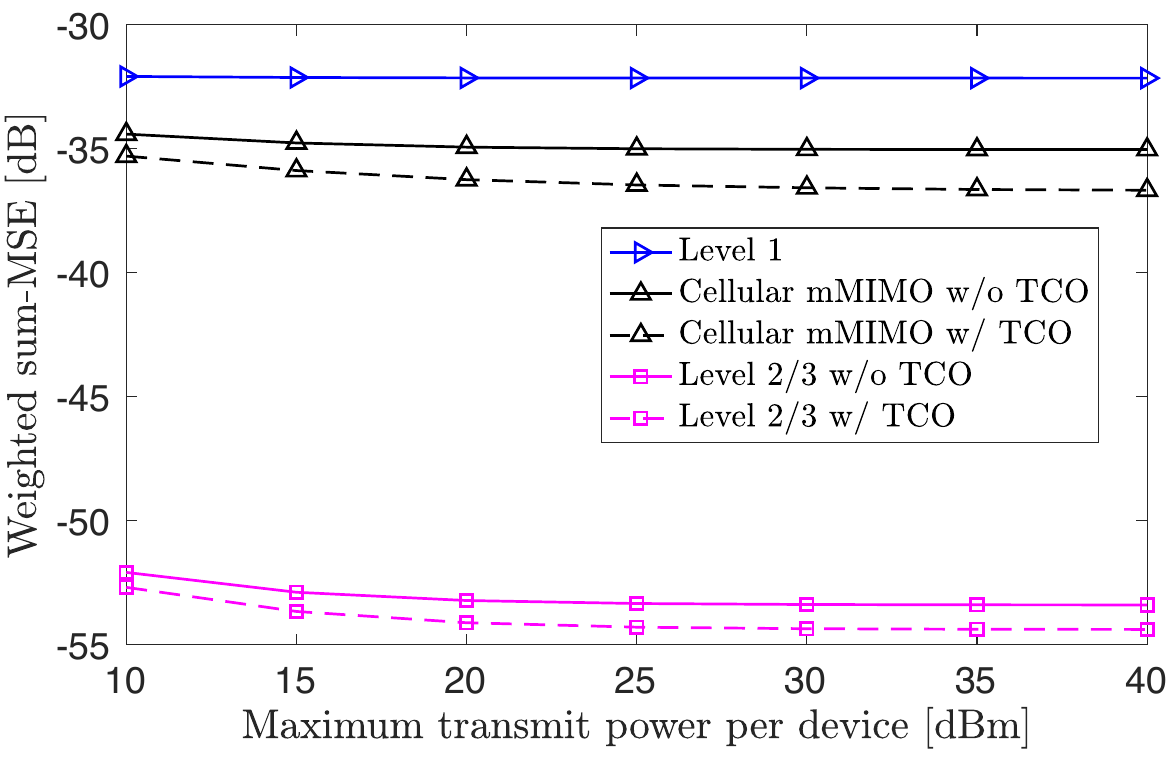}}
    \caption{MSE versus the maximum transmit power per device with $M=64$, $L=64$, $N=4$, $K=30$ and $G=3$.}
    \label{fig:G3}
\end{figure}

In Fig.~\ref{fig:G3}, we consider three FL groups and show the weighted sum-MSE of OtA model aggregation versus the transmit power budget per FL device. The general trend of the curves in Fig.~\ref{fig:G3} is the same as that in Fig.~\ref{fig:G1}. 
We can see that TCO brings an obvious performance gain in terms of the weighted sum-MSE, demonstrating the benefit of TCO in the presence of inter-group interference.
 Level~2/3 can outperform Cellular mMIMO in terms of the weighted sum-MSE by a large margin. However, the weighted sum-MSE of Level~1 is still  higher than that of Cellular mMIMO. Hence, Level~1 should not be adopted regardless of the number of FL groups and device distribution modes. 

\subsection{FL Test Accuracy}
Our analysis of FL convergence in Section \ref{sec:model_aggregation} is based on the assumption of a convex loss function. 
The loss function of a FNN with respect to the model parameter vector is generally non-convex~\cite{choromanska2015loss}.
In this subsection, we will show that the proposed OtA model aggregation designs are also applicable to FL using FNNs. 
We compare Cell-free mMIMO and Cellular mMIMO in term of FL convergence performance.
We also show the FL convergence performance for the error-free channel without channel fading and noise.
The results are the average of 10 independent FL trainings.

\begin{figure}[t]
    \centering
    \subfloat[Device Distribution Mode 1.]{\includegraphics[width=3.3in]{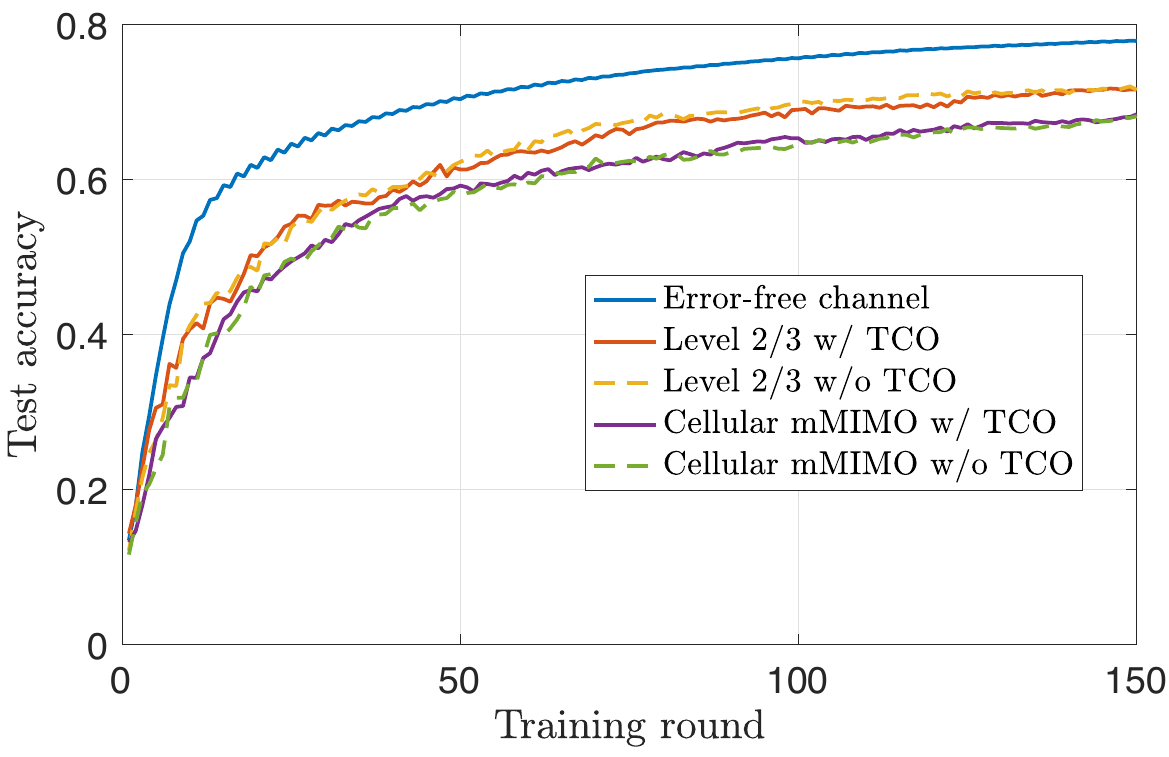}}
      \\
      \subfloat[Device Distribution Mode 2.]{\includegraphics[width=3.3in]{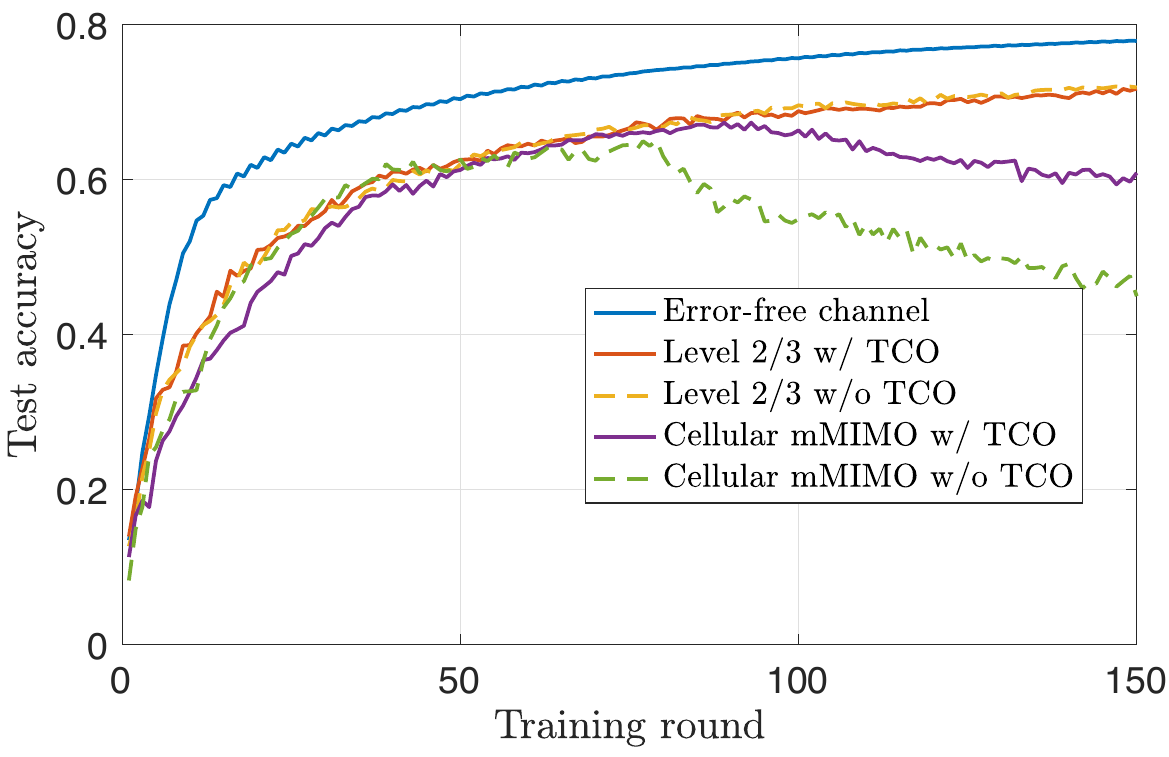}}
    \caption{FL test accuracy with $M=64$, $L=64$, $N=4$, $K=10$, $G=1$ and $P_{\mathrm{max}}=20$ dBm.}
    \label{fig:FL_G1}
\end{figure}

In Fig.~\ref{fig:FL_G1}, we consider one FL group and show the test accuracy versus the training iterations.   Comparing Fig.~\ref{fig:FL_G1} and Fig.~\ref{fig:G1}, we observe that a smaller MSE generally leads to a higher test accuracy.  We conclude that Level~2/3 achieves a good test accuracy under both device distribution modes. However, Cellular mMIMO fails to converge to a good test accuracy under Device Distribution Mode 2 due to the long transmission distances and large distance variations. Moreover, we see that there is still a gap between Level~2/3 and the error-free channel due to  OtA model aggregation errors. 

\begin{figure*}[t]
    \centering
    \subfloat[Group 1 (Fashion-MNIST).]{\includegraphics[width=2.15in]{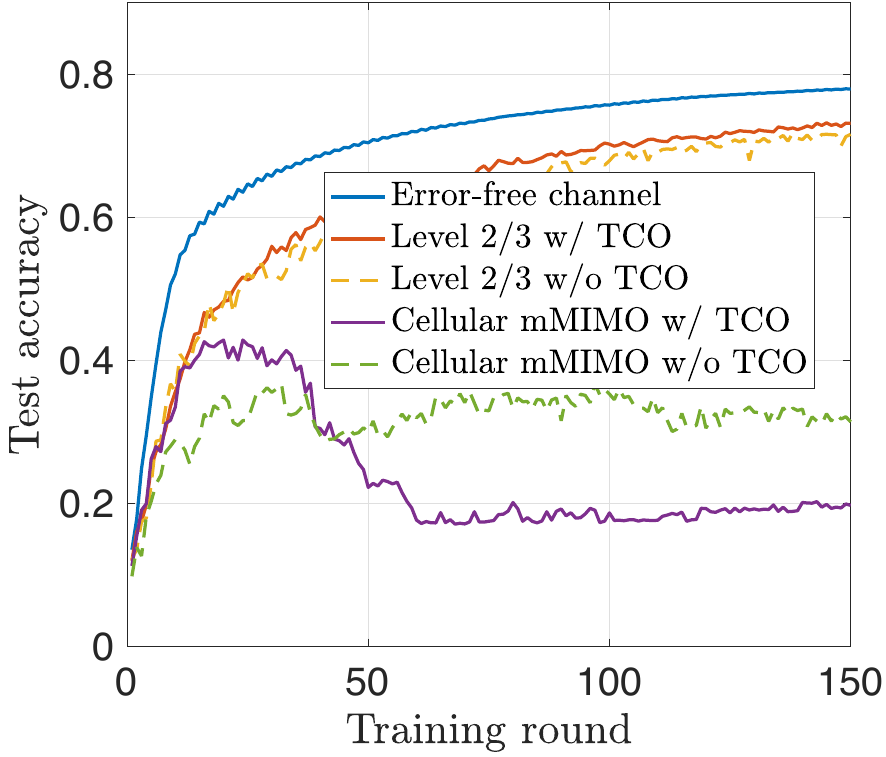}}
      \subfloat[Group 2 (MNIST).]{\includegraphics[width=2.15in]{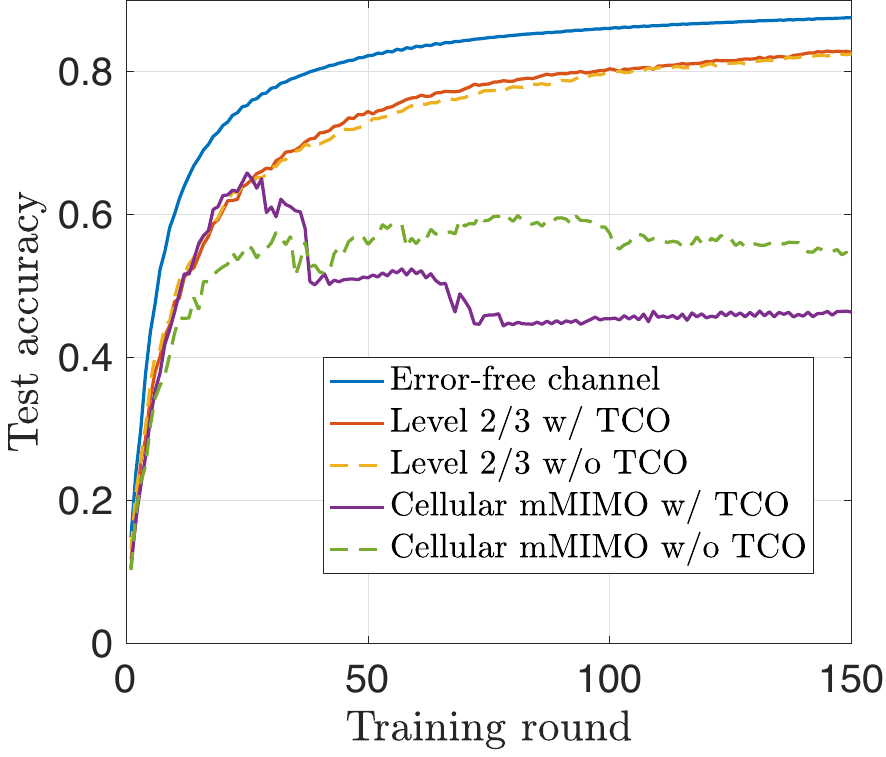}}
      \subfloat[Group 3 (EMNIST).]{\includegraphics[width=2.15in]{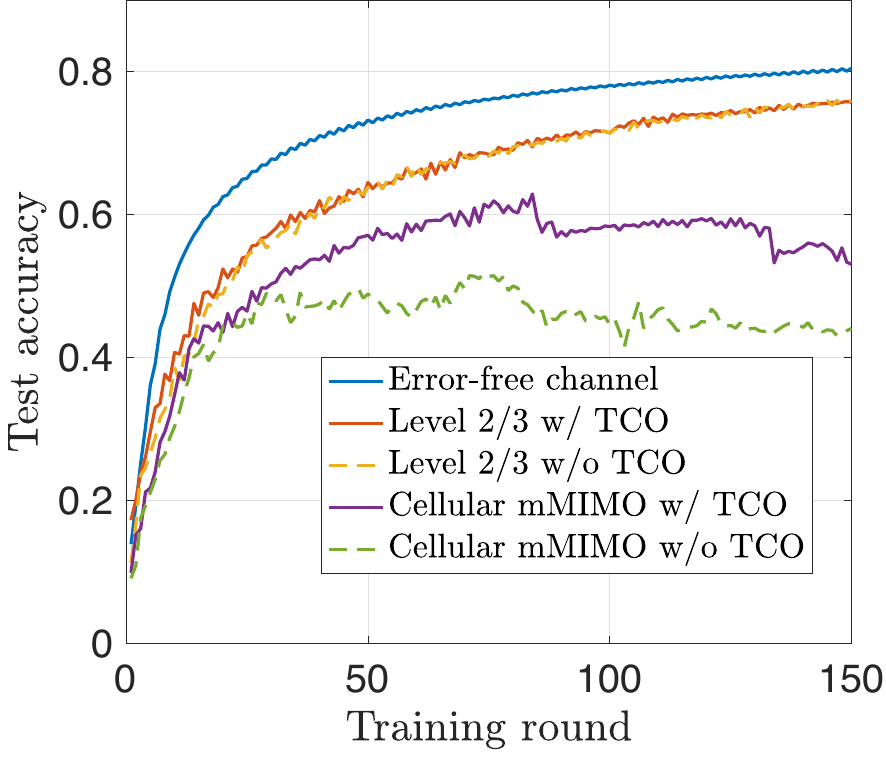}}
    \caption{FL test accuracy under Device Distribution Mode 1 with $M=64$, $L=64$, $N=4$, $K=30$, $G=3$ and $P_{\mathrm{max}}=20$ dBm.}
    \label{fig:FL_G3_DDM1}
\end{figure*}

\begin{figure*}[t]
    \centering
    \subfloat[Group 1 (Fashion-MNIST).]{\includegraphics[width=2.15in]{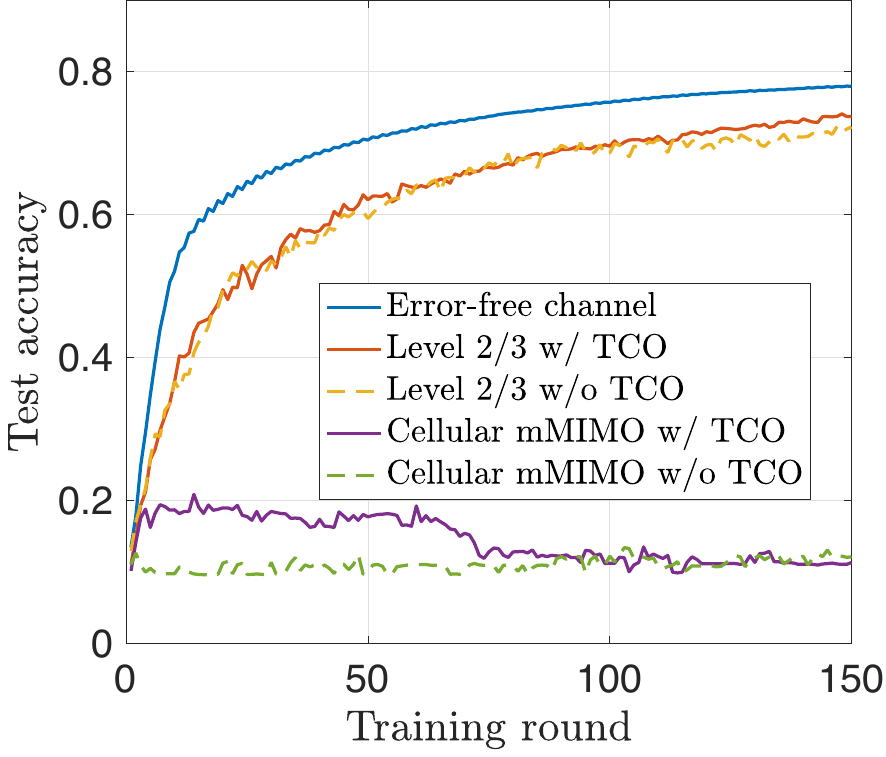}}
      \subfloat[Group 2 (MNIST).]{\includegraphics[width=2.15in]{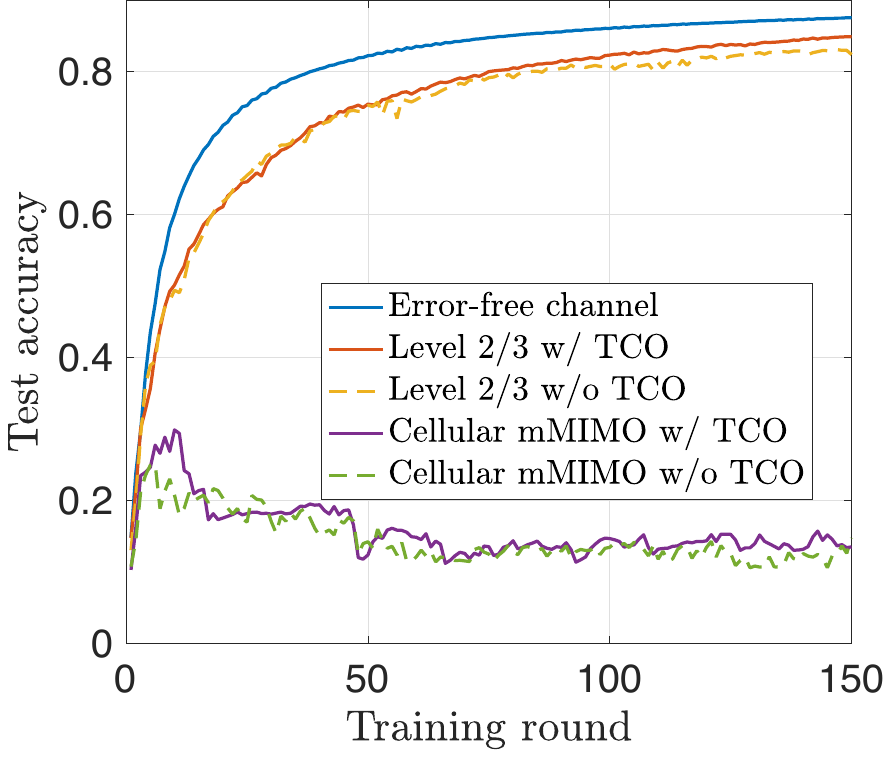}}
      \subfloat[Group 3 (EMNIST).]{\includegraphics[width=2.15in]{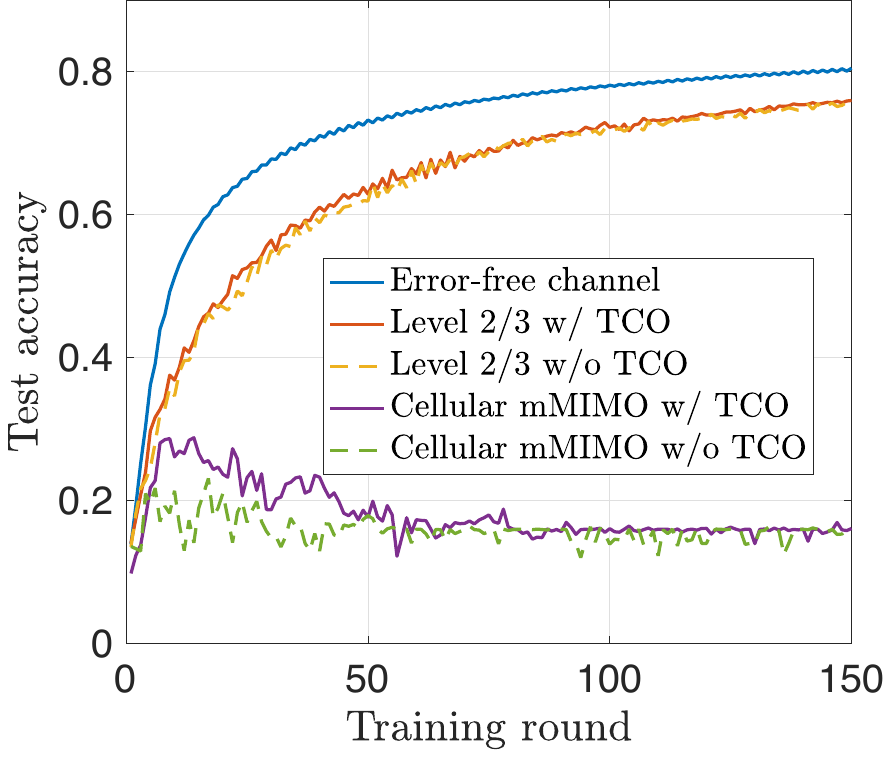}}
    \caption{FL test accuracy under Device Distribution Mode 2 with $M=64$, $L=64$, $N=4$, $K=30$, $G=3$ and $P_{\mathrm{max}}=20$ dBm.}
    \label{fig:FL_G3_DDM2}
\end{figure*}

In Fig.~\ref{fig:FL_G3_DDM1} and Fig.~\ref{fig:FL_G3_DDM2}, we consider 3 FL groups and present their convergence performance. We see that Level~2/3 has a good convergence performance in each FL task under both device distribution modes, indicating that Level~2/3 can well support multi-task OtA  FL. 
On the other hand, Cellular mMIMO cannot well address inter-group interference and fails to achieve a good convergence performance in each FL task under both device distribution modes.

\subsection{Comparison of Fronthaul Signaling}
Level~3 and Level~2 use the same combining vectors but have different fronthaul signaling requirements, as listed in Table~\ref{tab2}. This is because each AP at Level~2 converts the $N$-dimensional data signals into $G$-dimensional local estimates. Moreover, well-designed combining vectors need to be sent from the CPU to the APs. When $N<G$, it is clear that Level~2 requires more fronthaul signaling than Level~3. Assume that a coherence block includes $C$ training rounds. 
When $N>G$, Level~2 requires more fronthaul signaling than Level~3 if $\tau_u NL< \tau_u GL + CGNL$. After some simple calculations, the inequality becomes $\textstyle C > \frac{\tau_u(N-G)}{NG}$. The inequality holds when a coherence block includes a large number of training rounds. This happens in use cases of small AI models with a small number of model parameters.
In summary, Level~2 is more fronthaul friendly than Level 3 if $N>G$ and $\textstyle C < \frac{\tau_u(N-G)}{NG}$; otherwise, 
Level~3 is more fronthaul friendly. 
Level~1 always requires less fronthaul signaling than Level~2. However, it leads to a much higher MSE, as shown in Section~\ref{sec:MSE}. 


 \section{Conclusions}
\label{sec:conclusion}
In this paper, we  have studied multi-task OtA  FL in Cell-free mMIMO systems for three levels of AP cooperation. We have developed optimal OtA model aggregation designs for the considered AP cooperation levels.
Simulation results show that Cell-free mMIMO can significantly outperform Cellular mMIMO in terms of FL test accuracy by using 
fully centralized processing at Level~3 or partially distributed processing at Level~2.
Cell-free mMIMO operated at Level~3 and Level~2 can support multi-task OtA  FL regardless of device distribution. On the other hand, Cellular mMIMO fail to achieve good convergence performance when there are multiple FL groups and/or the FL devices are distributed across multiple cells.
Moreover, we have analyzed the fronthaul signaling overhead of different AP cooperation levels. We have pointed out the conditions under which Level~3 requires more fronthaul signaling than Level~2.
In the future, it will be of interest to investigate synchronization issues of OtA  FL in Cell-free mMIMO systems. In addition, we will significantly
extend the proposed approach to digital OtA  FL~\cite{razavikia2023channelcomp}.

\appendices
{
\section{Proof of Theorem \ref{theorem:FLconverge}}
\label{Appendix1}
Plugging (\ref{eq:varphikt}) into (\ref{eq:rgt}), we obtain that 
\begin{align}
\label{eq:thetag}
\widetilde{\bm{\theta}}_{g}^{t+1}=\sum_{k\in \mathcal{K}_{g}}\gamma_{kg}\left(\widehat{\bm{\theta}}_{g}^{t} - \eta_{g}\nabla F_{k}(\widehat{\bm{\theta}}_{g}^{t};\mathcal{D}_{k})\right).
\end{align}
From (2), we have 
\begin{align}
\label{eq:Fgtilde}
\nabla\widetilde{F}_{g}(\widehat{\bm{\theta}}_{g}^{t})=\sum_{k \in \mathcal{K}_{g}} \gamma_{kg}\nabla F_{k}(\widehat{\bm{\theta}}_{g}^{t};\mathcal{D}_{k}).
\end{align}
Plugging (\ref{eq:Fgtilde}) and $\textstyle\sum_{k\in \mathcal{K}_{g}}\gamma_{kg}=1$ into (\ref{eq:thetag}), we have
\begin{align}
\label{eq:thetag2}
\widetilde{\bm{\theta}}_{g}^{t+1}
=\widehat{\bm{\theta}}_{g}^{t}-
  \eta_{g}\nabla \widetilde{F}_{g}(\widehat{\bm{\theta}}_{g}^{t}).
\end{align}
Defining that $\mathbf{e}_{g}^{t}\triangleq \widetilde{\bm{\theta}}_{g}^{t} - \widehat{\bm{\theta}}_{g}^{t}$, (\ref{eq:thetag2}) can be rewritten as
\begin{align}
\label{eq:thetag3}
\widehat{\bm{\theta}}_{g}^{t+1}-\widehat{\bm{\theta}}_{g}^{t}
=-\mathbf{e}_{g}^{t+1} - 
  \eta_{g}\nabla \widetilde{F}_{g}(\widehat{\bm{\theta}}_{g}^{t}).
\end{align}

By plugging (\ref{eq:thetag3}) into (\ref{eq:Lipschitz}), we obtain that
\begin{align}
\label{eq:diff}
   &\widetilde{F}_{g}(\widehat{\bm{\theta}}_{g}^{t+1})- \widetilde{F}_{g}(\widehat{\bm{\theta}}_{g}^{t})   \nonumber \\ 
   &\le\left(\widehat{\bm{\theta}}_{g}^{t+1}-\widehat{\bm{\theta}}_{g}^{t}\right)^{\mathrm{T}}\nabla\widetilde{F}_{g}(\widehat{\bm{\theta}}^{t}_{g}) + \frac{\chi_{g}}{2}\left\|\widehat{\bm{\theta}}_{g}^{t+1}-\widehat{\bm{\theta}}_{g}^{t}\right\|^{2}\nonumber \\ 
   &=
   \left(-\mathbf{e}_{g}^{t+1} - 
  \eta_{g}\nabla \widetilde{F}_{g}(\widehat{\bm{\theta}}_{g}^{t})\right)^{\mathrm{T}}\nabla\widetilde{F}_{g}(\widehat{\bm{\theta}}^{t}_{g}) \nonumber \\
  &\quad \quad \quad \quad \quad \quad \quad \quad \quad \quad + \frac{\chi_{g}}{2}\left\|\mathbf{e}_{g}^{t+1} +
  \eta_{g}\nabla \widetilde{F}_{g}(\widehat{\bm{\theta}}_{g}^{t})\right\|^{2}
  \nonumber \\ 
   &=
   \left(-\mathbf{e}_{g}^{t+1} - 
  \eta_{g}\nabla \widetilde{F}_{g}(\widehat{\bm{\theta}}_{g}^{t})\right)^{\mathrm{T}}\nabla\widetilde{F}_{g}(\widehat{\bm{\theta}}^{t}_{g})+ \nonumber \\
  &\ \frac{\chi_{g}}{2}\left\|\mathbf{e}_{g}^{t+1}\right\|^{2}+
  \chi_{g}\eta_{g}(\mathbf{e}_{g}^{t+1})^{\mathrm{T}}\nabla \widetilde{F}_{g}(\widehat{\bm{\theta}}_{g}^{t})+\frac{\chi_{g}\eta_{g}^{2}}{2}\left\|\nabla \widetilde{F}_{g}(\widehat{\bm{\theta}}_{g}^{t})\right\|^{2}.
\end{align}
Setting $\textstyle\eta_{g}=\frac{1}{\chi_{g}}$ in (\ref{eq:diff}), we obtain that 
\begin{align}
\label{eq:57}
   \widetilde{F}_{g}(\widehat{\bm{\theta}}_{g}^{t+1})- \widetilde{F}_{g}(\widehat{\bm{\theta}}_{g}^{t})   \le \frac{\chi_{g}}{2}\left\|\mathbf{e}_{g}^{t+1}\right\|^{2}-
   \frac{1}{2\chi_{g}}\left\|\nabla \widetilde{F}_{g}(\widehat{\bm{\theta}}_{g}^{t})\right\|^{2}.
\end{align}
From \cite{friedlander2012hybrid}, we have 
\begin{align}
\textstyle\left\|\nabla \widetilde{F}_{g}(\widehat{\bm{\theta}}_{g}^{t})\right\|^{2}\!\ge\! 2\xi_{g}\left(\widetilde{F}_{g}(\widehat{\bm{\theta}}_{g}^{t})-\widetilde{F}_{g}(\widetilde{\bm{\theta}}_{g}^{\mathrm{opt}})\right). 
\end{align}
Accordingly, (\ref{eq:57}) is further derived as
\begin{align}
\label{eq:59}
   &\widetilde{F}_{g}(\widehat{\bm{\theta}}_{g}^{t+1})- \widetilde{F}_{g}(\widehat{\bm{\theta}}_{g}^{t})   \le \nonumber \\ 
   &\quad \quad \quad \quad \quad  \frac{\chi_{g}}{2}\left\|\mathbf{e}_{g}^{t+1}\right\|^{2}-
   \frac{\xi_{g}}{\chi_{g}}\left(\widetilde{F}_{g}(\widehat{\bm{\theta}}_{g}^{t})-\widetilde{F}_{g}(\widetilde{\bm{\theta}}_{g}^{\mathrm{opt}})\right).
\end{align}
Subtracting $\widetilde{F}_{g}(\widetilde{\bm{\theta}}_{g}^{\mathrm{opt}})$ and taking the expectations on both sides of (\ref{eq:59}), we have
\begin{align}
\label{eq:60}
   &\mathbb{E}\left\{\widetilde{F}_{g}(\widehat{\bm{\theta}}_{g}^{t+1})- \widetilde{F}_{g}(\widetilde{\bm{\theta}}_{g}^{\mathrm{opt}})\right\}   \le \nonumber \\ 
   &\ 
   \left(1-
   \frac{\xi_{g}}{\chi_{g}} \right)\mathbb{E}\left\{\widetilde{F}_{g}(\widehat{\bm{\theta}}_{g}^{t})-\widetilde{F}_{g}(\widetilde{\bm{\theta}}_{g}^{\mathrm{opt}})\right\} + \frac{\chi_{g}}{2}\mathbb{E}\left\{\left\|\mathbf{e}_{g}^{t+1}\right\|^{2}\right\}.
\end{align}
Finally, by recursively applying (\ref{eq:60}) for $T$ times, we obtain (\ref{eq:T}), which completes the proof.
}

\normalem
\bibliographystyle{IEEEtran}



\end{document}